\documentclass{aa}
\usepackage[varg]{txfonts}
\usepackage[english]{babel}
\usepackage{amsmath}
\usepackage{graphicx}
\usepackage{subcaption}
\usepackage{natbib}
\bibpunct{(}{)}{;}{a}{}{,} % to follow the A&A style

\graphicspath{{.}}

\begin{document}
\title{The ALMA Lupus protoplanetary disk survey: evidence for compact gas disks and molecular rings from CN}
\titlerunning{The ALMA Lupus protoplanetary disk survey: CN}

\author{S.E.~van Terwisga\inst{\ref{leiden1}}\and E.F. van Dishoeck\inst{\ref{leiden1}\and\ref{garching}}\and P. Cazzoletti\inst{\ref{garching}}\and S. Facchini\inst{\ref{ESO}}\and L. Trapman\inst{\ref{leiden1}}\and J.P. Williams\inst{\ref{hawaii}}\and C.F. Manara\inst{\ref{ESO}}\and A. Miotello\inst{\ref{ESO}}\and N. van der Marel\inst{\ref{hawaii}\and\ref{victoria}}\and M. Ansdell\inst{\ref{hawaii}\and\ref{berkeley}}\and M.R. Hogerheijde\inst{\ref{leiden1}\and\ref{api}}\and M. Tazzari\inst{\ref{cambridge}}\and L. Testi\inst{\ref{ESO}\and\ref{INAF}}}

\institute{
Leiden Observatory, Leiden University, PO Box 9513, 2300 RA Leiden, The Netherlands\label{leiden1}\\ \email{terwisga@strw.leidenuniv.nl}\and
Max-Planck-Institut f{\"u}r Extraterrestrische Physik, Gie{\ss}enbachstraße, D-85741 Garching bei M{\"u}nchen, Germany\label{garching} \and
Institute for Astronomy, University of Hawai`i at M{\=a}noa, 2680 Woodlawn Dr., Honolulu, HI, USA\label{hawaii} \and
Herzberg Astronomy \& Astrophysics Programs, NRC of Canada, 5017 West Saanich Road, Victoria, BC V9E 2E7, Canada\label{victoria} \and
Department of Astronomy, University of California, Berkeley, CA 94720, USA\label{berkeley} \and
Institute of Astronomy, University of Cambridge, Madingley Road, CB3 0HA, Cambridge, UK\label{cambridge} \and
Anton Pannekoek Institute for Astronomy, University of Amsterdam, Postbus 94249, 1090 GE Amsterdam, The Netherlands\label{api} \and
European Southern Observatory, Karl-Schwarzschild-Str. 2, D-85748 Garching bei M{\"u}nchen, Germany\label{ESO} \and
INAF-Osservatorio Astrofisico di Arcetri, Largo E. Fermi 5, I-50125 Firenze, Italy\label{INAF}
}

\abstract
{The cyanide radical CN is abundant in protoplanetary disks, with line fluxes often comparable to those of $^{13}$CO. It is known to be sensitive to UV irradiation of the upper disk atmosphere, with models predicting ring-shaped emission.}
{We seek to characterize the CN emission from 94 Class II disks in the Lupus star forming region, compare it to observations in other regions, and interpret our observations with a grid of models. The CN emission morphology is discussed for two primordial disks, Sz 71 and Sz 98, and modeled in more detail.}
{ALMA observed CN $N=3-2$ in Lupus disks down to sensitivities better than previous surveys. Models constructed with the physico-chemical code~\texttt{DALI} are used to study the integrated fluxes of the disks and resolved emission of CN in disks without (dust) substructures.}
{CN $N=3-2$ is bright, and detected in $38\%$ of sources, but its disk-integrated flux is not strongly correlated to either $^{13}$CO or continuum flux. Compared to pre-ALMA single-dish surveys, no significant difference in the CN flux distributions in Lupus and Taurus-Auriga is found, although $\rho$ Ophiuchus disks may be fainter on average. We find ring-shaped CN emission with peak radii of $\sim 50$\,AU in two resolved disks.}
{A large fraction of sources is faint in CN; only exponential gas surface density cutoffs at $R_{\rm{c}} \leq 15$\,AU can reconcile observations with models. This is the first observational evidence of such a compact gas disk population in Lupus. Absolute intensities and the emission morphology of CN are reproduced by \texttt{DALI} models without the need for any continuum substructure; they are unrelated to the CO snowline location. These observations and the successful modeling of these rings provide a new probe of the structure and conditions in disks, and particularly their incident UV radiation field, if disk size is determined by the data.}


\maketitle

\section{Introduction}
	The primary motivation for the study of protoplanetary disks is to understand the formation of planetary systems. For this purpose it is essential to characterize the structure and composition of both dust and gas in the disks, with different molecules providing different information on the gas structure and conditions.
	
	Initially, most disk observations were limited by low spatial resolution and sensitivity, and focused on disk-integrated observations of well-known, bright objects~(see the review by \citealt{henning13}). In recent years, ALMA has been instrumental in detecting as well as resolving the mm-sized dust and a number of gas species near the midplanes of protoplanetary disks, thus providing a new window on the environment where planets are formed. A number of large, unbiased, and deep surveys have significantly improved our understanding of the protoplanetary disk populations in nearby star-forming regions \citep[e.g.][]{andrews05, pascucci16, barenfeld16, ansdell16}. Among the puzzling results from these surveys is that the CO isotopologue emission is weaker than expected when assuming normal CO abundances and an ISM-like gas-to-dust ratio \citep{ansdell16,miotello17,long17}.

	While most of these surveys focused exclusively on the (resolved) continuum emission of Class II disks, or on CO and its isotopologues, relatively few surveys have targeted other molecular species. Targeting other molecules is scientifically valuable: if their chemistry is sufficiently well understood, they can provide additional independent constraints on the properties of disks. However, molecular lines are often faint, or the chemical pathways behind their formation and destruction not well understood, making them less attractive targets for a survey.
	
	The cyanide radical (CN) is, in this context, of particular interest. After $^{12}$CO, it is one of the brightest molecules in disks, and has line fluxes comparable to or higher than those of $^{13}$CO \citep[e.g.][]{dutrey97,thi04,oeberg11,salter11}. Due to its brightness, several large single-dish surveys of this molecule have already been performed, using the IRAM 30-meter telescope \citep{guilloteau13,reboussin15}. Together these surveys cover 74 disks in Taurus and $\rho$ Ophiuchi. However, the samples studied suffer from two key drawbacks: they are not complete, and are biased towards the brightest or most radially extended sources at millimeter wavelengths.
	
	Despite these issues, the abundance of CN has been suggested to be enhanced in Class II versus Class I disks \citep{kastner14,guilloteau16}. Also, \citet{guilloteau14} have successfully used the cyanide radical as a stellar mass probe. However, in order to get a better understanding of the dependence of CN on disk properties, it is necessary to use an unbiased, large, and uniformly observed sample of disk observations.
	
	The observational basis for this study is the complete survey of Class II disks in the Lupus star-forming region \citep{ansdell16,ansdell18}. This survey offers good angular resolution at $\sim0.3''$, or a $\sim 24$\,AU radius at $160$\,pc (the typical distance for these objects, based on {\it Gaia} DR2 \citealp{gaiadr2,bailerjones18}). The sensitivity of this dataset is a factor of $\sim 3$ better than previous large single-dish CN surveys. Moreover, the sample studied here is unbiased. By covering 94 of 95 Class II disks in Lupus I, III, and IV with how star masses $\geq 0.1\,M_{\odot}$, this survey is very complete and more than doubles the total number of disks observed in CN.
	
	From an evolutionary perspective, the objects in the Lupus survey must be relatively close in age, around $1-3$\,Myr \citep{sfr_book}, and formed in a similar environment. This reduces the impact of different disk ages on observational properties. Finally, the Lupus disks have been characterized at multiple wavelengths. Apart from the dust continuum, the Lupus disk survey targeted $^{12}$CO and its most common isotopologues $^{13}$CO and C$^{18}$O. Additionally, intermediate-resolution {\it XSHOOTER} spectra of most sources in the sample have also been taken \citep{alcala14,alcala17}, so that accurate measurements of the host star properties and accretion rates are available.
	
	 Apart from providing disk-integrated observations of CN towards many sources, resolved images are also obtained, even at integration times of just one minute. In comparison to pre-ALMA interferometric studies of individual objects  \citep[e.g.][]{oeberg11, guilloteau14}, our effective physical resolution is improved by a factor of $\geq 2$. With these short integration times, the $S/N$ of the molecular gas emission is low: for the brightest sources in CN, the line peak has an $S/N \sim 12$ relative to the channel noise. However, we are able to discuss in greater detail the ring-like radial distribution of CN emission in two favorably inclined, bright sources, and study its underlying causes in a general way.
	 
	 This is relevant because, in recent years, the high-resolution capabilities of ALMA have lead to the discovery of many rings and other radial structures, in gas emission \citep[e.g.][]{kastner15,teague17} as well as in dust \citep[e.g.][]{HLTau, isella16}, in a variety of sources and molecules. These structures occur both in transitional disks, which are characterized by central cavities in their dust distribution \citep{espaillat14,nienke16a} as well as in full disks, which are the focus of this article. Indeed, ALMA observations of TW Hya also show CN to be distributed in a ring \citep{teague16}, suggesting this type of emission morphology may be more common for this species.
	 
	 For molecular species with ring-like emission, several different processes have been identified as their root cause: CO is depleted in at least some of the gaps in the dust in the HD 163296 system \citep{isella16}, while CS emission coincides with a region of decreased dust surface density in TW Hya \citep{teague17}. More complex chemical processes also play a role for some species: dust grains locking up volatile carbon and oxygen in the outer disk have been suggested as an explanation for rings observed in C$_2$H and $c$-C$_3$H$_2$ \citep{kastner15,bergin16}. Snow lines are also an important mechanism for causing the ring-like emission of some molecules. For instance, N$_2$H$^+$ is distributed in a ring in several disks \citep{qi13,qi15}, of which the radius is sensitive to the position of the CO ice line \citep{aikawa15,merel17}. Many other molecular species are now showing a rich variety of emission morphologies, including single and multiple rings: DCO$^{+}$, DCN, H$^{13}$CN, and H$^{13}$CO$^+$, H$_2$CO \citep{mathews13,oberg15,huang17,carney17,salinas17}. However, for many of these species, finding the underlying cause of the emission morphology from observations alone is difficult, due to the variety in emission profiles between different molecules, and between disks.
	
	To study integrated fluxes and the radial distribution of CN emission in disks, the chemical network that sets its formation and destruction must be understood. Significant advances have recently been made on this front: the physical-chemical code~\texttt{DALI} \citep{dali1,dali2}, which self-consistently calculates the chemical and thermal structure of the disk models, has been expanded to include nitrogen chemistry \citep{visser18} using the most recent rates and branching ratios from \citet{loison14}. Moreover, \texttt{DALI} now uses updated UV cross-sections for CN photodissociation \citep{alan17}. This allows us to predict the CN emission expected for a wide variety of disks, and study how the models depend on various disk parameters, such as radius, mass, and incident UV flux \citep{cazzoletti18}. Several key predictions have been found: first, an increasing integrated CN flux with increasing disk size; second, a strong dependence of CN flux on the UV radiation field of stars; and a relatively weak dependence on total disk mass and on the volatile carbon abundance. In terms of the distribution of CN emission over the disk, a ring-like morphology is always found in the models, with the ring radius depending on the characteristic radius $R_{\rm{c}}$ of the gas disk, the vertical profile of the disk, and the UV radiation field.

	In this paper, we discuss the CN emission in the full Lupus Class-II disk population (Section~\ref{sec:cnpopres}), investigate possible correlations with other disk tracers, and compare the Lupus disks to previous large single-dish surveys. In Section~\ref{sec:cnvspop}, the properties of the full disk sample are compared to a \texttt{DALI} model grid, focusing on the consequences for disk gas radii and the sensitivity to UV radiation. To investigate resolved CN emission, two full (i.e. non-transitional) disks are described in detail (Section~\ref{sec:twodisksimage}). These disks -- Sz 98 and Sz 71 -- are amongst the brightest sources in CN, have clearly detected and resolved emission in both lines and continuum, and are not strongly inclined (around $45^{\circ}$). Section~\ref{sec:dalianalysis} discusses individual \texttt{DALI} models for CN emission in Sz 98 and Sz 71, to see if a smooth power-law surface density distribution can reproduce their CN intensity profiles.

\section{Observations and data reduction}
\label{sec:obs}
The data for this project are part of the ALMA Lupus disk survey \citep{ansdell16,ansdell18}, which targeted 86 Class II stars with ALMA Band 7 and 6 (ID: 2013.1.00220.S, 2015.1.00222.S, PI: J. Williams), combined with the 7 sources from the Lupus Completion Survey (ID: 2016.1.01239.S, PI: S. van Terwisga). To these 93 disks we add previously-published data for IM Lup \citep{oeberg11}. On top of the $^{12}$CO $J=2-1$, $^{13}$CO and C$^{18}$O $J=2-1$ and $J=3-2$ lines, and $335$~and $225$ GHz continuum, the spectral settings of these surveys  targeted the CN $\nu=0$ $N=3-2$ transition in Band 7. Of the fine-structure transitions of this molecule, we focus on the brightest ($J=7/2-5/2$) line, and in particular the $F=7/2-5/2$ and $F=9/2-7/2$ transitions at 340.247770\,GHz and the partially overlapping $F = 5/2 - 3/2$ transition; all others are too faint to be detected in our data but these transitions together are responsible for $95\%$ of the flux in the $J=7/2-5/2$ line in the optically thin case~\citep{hilyblant17}. CN was observed at a lower spectral resolution of 0.24\,MHz compared to $^{13}$CO $J=3-2$, for which a 0.12\,MHz resolution was used. The analysis in this article mainly focuses on the Band 7 results, and only uses the Band 6 $^{12}$CO $J=2-1$ data. 

	In Band 7, all M-type stars were observed for 1 minute on source, while earlier-type stars were observed for 30\,s, giving continuum sensitivities of $0.25$ and $0.41$\,mJy beam$^{-1}$. Band 6 integration times were twice as long. For the sources in the Lupus Completion Survey longer integration times were used, of 3.5 and 4.5 minutes in Band 6 and Band 7 respectively, but a smaller beam lead to similar sensitivities \citep{ansdell18}; otherwise all observing parameters were identical.
	
	Pipeline calibration of the data was performed by the NRAO. Flux, phase, bandpass and gain were calibrated using observations of Titan, J1427-4206, and J1604-4228 and J1610-3958 respectively. We also performed self-calibration on the data in order to maximize the $S/N$ of the resulting images. Phase- and gain self-calibration were performed for all objects at the largest solution interval of one minute (for the highest sensitivity). These disks are not so bright as to be dominated by phase- and gain noise even in the continuum images, but this procedure resulted in an improvement in $S/N$.
	
	Both continuum and line imaging were performed using the {\it clean} task in \texttt{CASA}. The continuum of the sources discussed in detail was imaged using Briggs weighting with a robust parameter of $+0.5$ after averaging over all continuum channels. This optimized our resolution and $S/N$ in the image. The resulting effective beam size in Band 7 is $0.3'' \times 0.3''$, while in Band 6, it is $0.27'' \times 0.27''$. In the Lupus completion survey, the beams were $0.25'' \times 0.22''$ (Band 6) and $0.19'' \times 0.18''$ (Band 7), respectively.

	The CN channels were binned for an effective velocity resolution of 0.5\,km\,s$^{-1}$, and imaged using natural weighting to maximize $S/N$. The resulting typical beam shape is $0.37'' \times 0.31''$; for the Lupus completion survey it is $0.22'' \times 0.20''$. The $^{12}$CO $J=2-1$ and $^{13}$CO $J=3-2$ lines were imaged with the same parameters.	In these observations, the typical resulting channel rms of these CN observations is 10\,mJy\,beam$^{-1}$. The median noise level on the disk- and velocity integrated line fluxes is 48 mJy\,km\,s$^{-1}$.
	
	Of the 93 disks observed with ALMA, we focus specifically on the two brightest disks that are moderately inclined ($<50^{\circ}$), resolved across multiple beams, have high $S/N$ moment-zero maps, and do not show any signs of depleted inner dust cavities. For these two disks, Sz 98 and Sz 71, it is possible to study the spatial distribution of CN. To generate and analyze the moment-zero maps at the highest sensitivity without introducing biases (by, e.g., using significance cuts in the individual channels), image-plane Keplerian masking was used to create their moment-zero maps \citep{loomis18,salinas17}. For this method to work, several properties of the system must be known. The stellar masses have been derived by determining the stellar properties of the targets with optical spectroscopy \citep{alcala17} and comparing the stellar temperature and luminosity with evolutionary models by \citet{siess00}, while the mm-continuum observations allowed us to constrain the disks' position angles and inclinations using the {\it imfit} task in \texttt{CASA}, version \texttt{4.3.1}. These results are combined with a line-width uncertainty factor (here taken as 0.5\,km\,s$^{-1}$ for CN, and 0.3\,km\,s$^{-1}$ for the CO isotopologue lines) which accounts for uncertainties in the precise stellar position, the disk position angle and inclination, and the effect of adjacent hyperfine lines in CN (specifically, the partial overlap of the $F=7/2-5/2$ and $F=9/2-7/2$ transition with the $F=5/2-3/2$ transition). The resulting mask allowed us to use only those voxels in the objects' data cubes in which a signal can be reasonably expected. Compared to a normal moment-zero map, the biggest gain in sensitivity using this mask is in the outer regions of the disk, where (often) only a single channel contains a line signal. For $^{13}$CO in Sz 98, we additionally used $u,v$-plane tapering to create an $0.5''$ circular beam, in order to maximize $S/N$.

\section{Integrated CN flux survey results}
\label{sec:cnpopres}

%% CN vs continuum & $^13$CO
\begin{figure*}[ht]
\begin{center}
		\includegraphics[width=\textwidth]{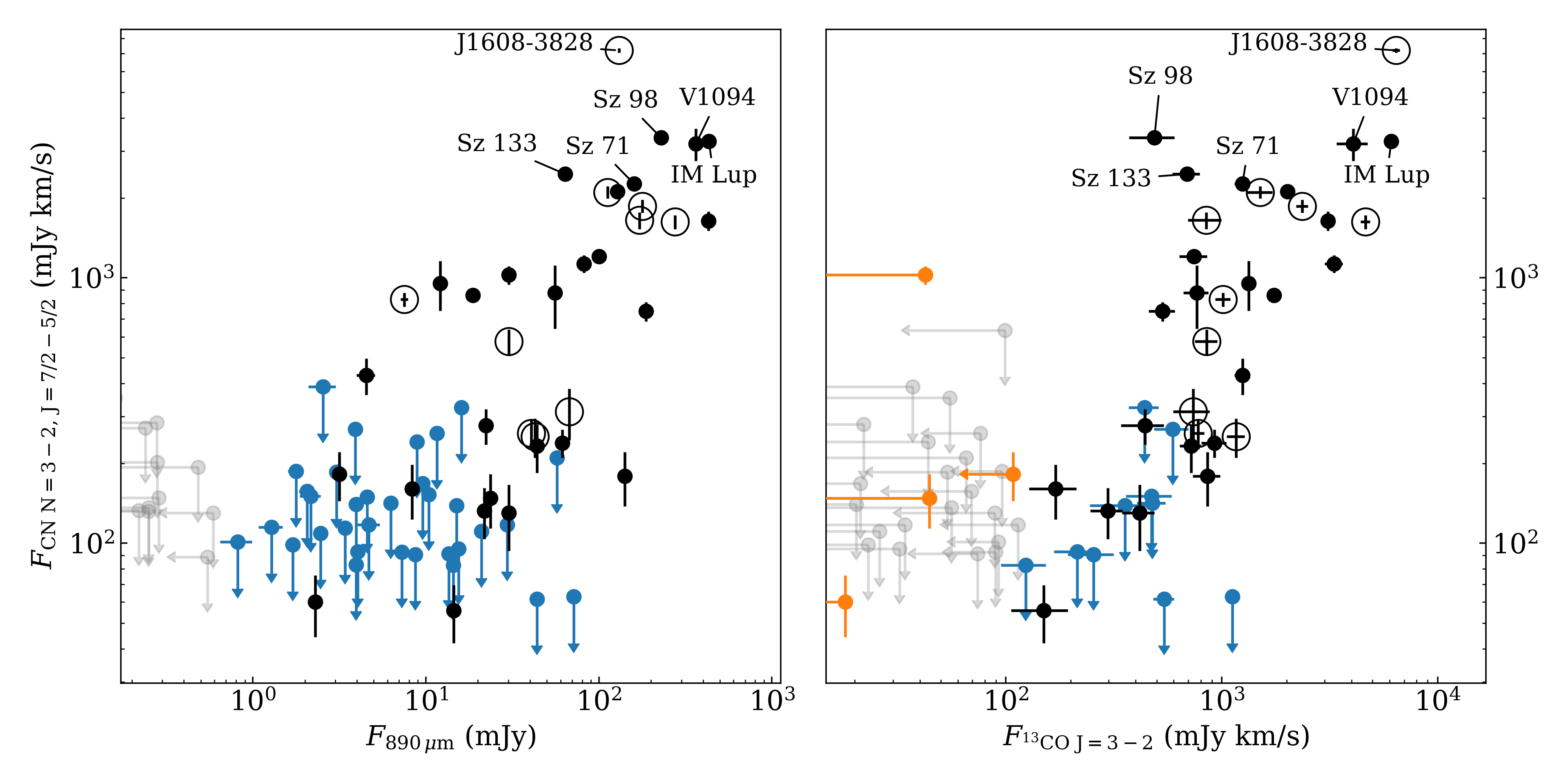}
		\caption{Disk-integrated CN $N=3-2,~J=7/2-5/2$ versus Band 7 continuum {\it(left panel)} and integrated $^{13}$CO $J=3-2$ fluxes {\it(right panel)} for the Lupus disk sample. Detected sources are black, CN upper limits are blue, $^{13}$CO upper limits orange, and double upper limits are gray. Transition disks are marked with a circle. All upper limits are at the $3\sigma$ level. The brightest 6 disks are labeled.}
		\label{fig:cnvscont}
\end{center}
\end{figure*}

	The full Lupus sample consists of 95 Class II disks, 93 of which have been observed in CN in this survey. Of the two missing objects, Sz 91 and IM Lup, only Sz 91 has no CN observations available. Instead of the CN $N=3-2$ transition, CN $N=2-1$ has been observed and detected towards IM Lup by \citet{oeberg11}.

	In these observations, 36 of 94 sources are detected, for a total detection rate of $38\%$ across the sample. All but 3 of these sources are also detected in $^{13}$CO. The brightest sources in terms of CN flux are the transition disk J16083070-3828268 \citep{vandermarel18}, Sz 98, V1094 Sco \citep{svt18}, IM Lup, Sz 133, and Sz 71. Interestingly, the transition disks with large inner cavities in the sample that were identified in \citet{vandermarel18} are all detected in CN. These sources were also found to be among the brightest in the sample in continuum and $^{13}$CO emission. Integrated fluxes for these objects were determined in the same way as those in \citet{ansdell16,ansdell18}: circular aperture photometry was performed on the source position after integrating over all channels with significant emission, with the smallest aperture containing the full disk flux determined using a curve-of-growth method. For non-detections, a beam-sized aperture was used, and a conservative velocity range with a width of $10$\,km\,s$^{-1}$, to be consistent with our previously used noise definitions. 

	\subsection{CN versus other disk tracers: $^{13}$CO and continuum}
	The large, unbiased sample of Class II disks in Lupus allows us to compare the resulting CN fluxes to the other important disk tracers: their continuum flux (proportional to the disk dust mass, if we assume the emission to be optically thin at these wavelengths) and integrated $^{13}$CO $J=3-2$ emission, which is the most commonly detected CO isotopologue, and more optically thin than $^{12}$CO $J=2-1$. The results of this comparison are presented in Figure~\ref{fig:cnvscont}. Appendix~\ref{app:othertracer} shows CN versus the C$^{18}$O emission in Figure~\ref{fig:cnvsc18o}, but we do not discuss it further here, due to the low detection rate of this molecule in both Band 6 and Band 7.
	
	The CN fluxes detected towards Lupus span several orders of magnitude, and are presented in full in Table~\ref{tab:all_cn}. Compared to the continuum- and $^{13}$CO fluxes, a large scatter becomes apparent, of several orders of magnitude, especially due to the impact of upper limits in the data. Consider, for example, the position of Sz 71 (labeled) in both panels of Figure~\ref{fig:cnvscont}: a source with similarly bright $^{13}$CO is undetected in CN, while another disk with a similar 890\,$\mu$m flux has CN emission that is an order of magnitude fainter. Similar behavior is seen in the other direction.
	
	To formally test for correlations between these observables, a modified Kendall-rank test taking into account upper limits was used \citep{isobe86}. This test confirms that neither the $890\,\mu$m continuum nor the $^{13}$CO $J=3-2$ line flux are correlated with the CN flux ($p > 0.05$), consistent with the results from the IRAM 30-m telescope survey in Taurus-Auriga by \citet{guilloteau13}. Not being able to reject the null result does not mean that no information can be extracted from this figure; however, to do so, more sophisticated models are needed, which are discussed in Section~\ref{sec:cnvspop}.

	\subsection{CN in Lupus versus Taurus-Auriga and $\rho$ Ophiuchi}
	Comparing the results of the Lupus sample to those of other literature surveys allows us to test if different regions (possibly with different ages, typical disk sizes, or external UV fields) have different CN emission properties. Here, we compare the Lupus CN observations to the results presented for CN in \citet{guilloteau13} and \citet{reboussin15}, who targeted 42 and 30 stars respectively with the IRAM 30-m single-dish telescope, primarily in the Taurus-Auriga and $\rho$ Oph star-forming regions.

\begin{figure}[ht]
\begin{center}
		\includegraphics[width=0.5\textwidth]{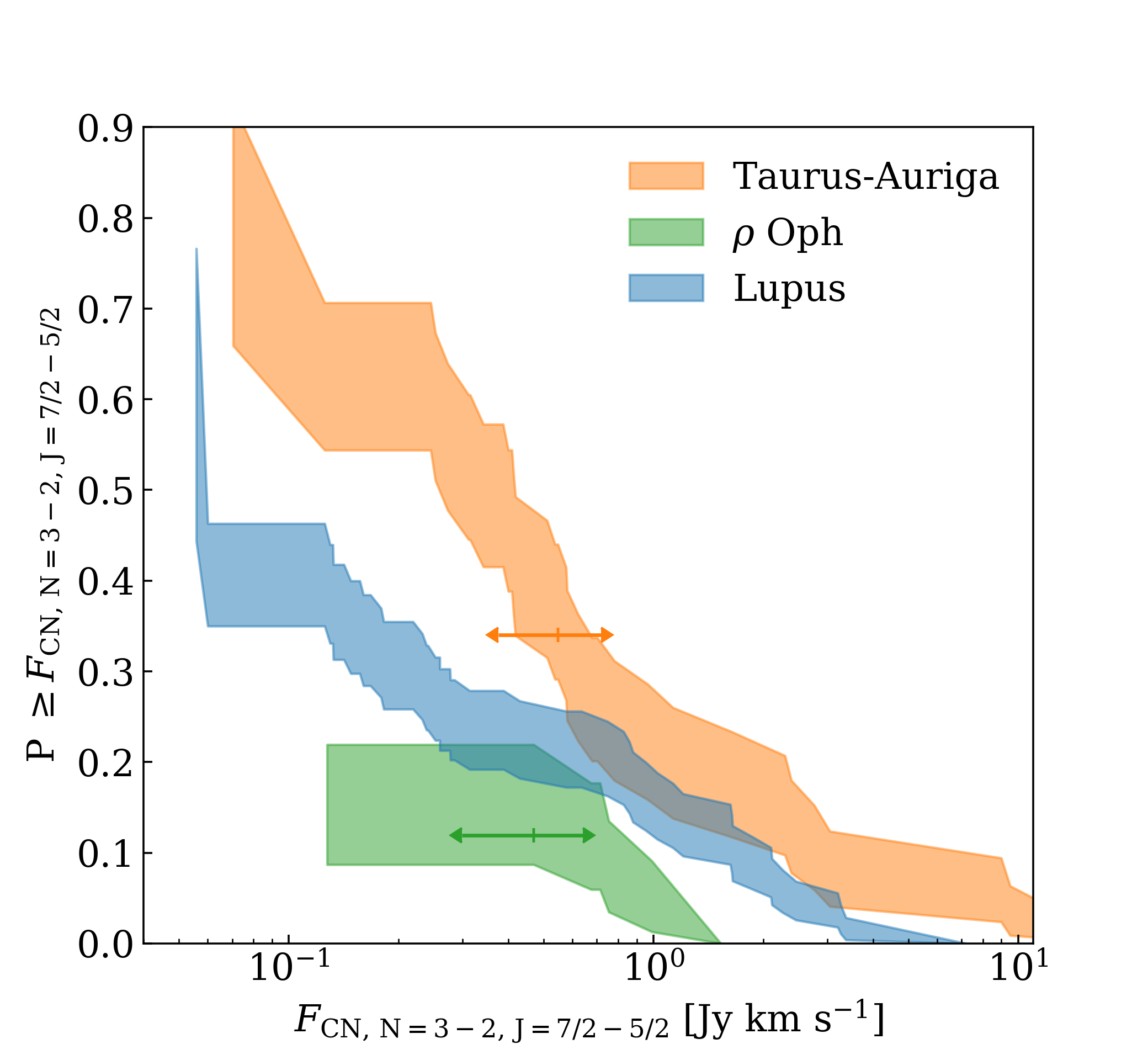}
		\caption{Disk-integrated CN $N=3-2,~J=7/2-5/2$ fluxes in the complete Lupus disk survey (this work) compared to those of a sample of disks in Taurus-Auriga \citep{guilloteau13} and $\rho$ Oph \citep{reboussin15}. Source distances for all objects were taken from {\it Gaia} DR2, if available; otherwise, median distances to the three populations were used \citep{gaiadr2,bailerjones18}. Fluxes were then scaled to the median distance of Lupus ($160$\,pc). Colored arrows indicate the effect of using different CN $(N=2-1~J=5/2-3/2)~/~( N=3-2~J=7/2-5/2)$ scaling parameters over the full range of model values.}
		\label{fig:cndistro}
\end{center}
\end{figure}	
	
	Some caveats apply to this analysis. First, the IRAM 30-m sample covers a range of spectral types, stellar luminosities, and disk masses, but might be biased to bright or radially extended targets. Large-scale cloud contamination is a possibility, but ruled out by the authors of both studies. Likewise, we must assume similar underlying stellar properties in all these samples. Fortunately, this assumption has been found to hold previously for these regions~\citep{ansdell16,cieza18}, and we do not expect this to have a significant effect.
	
	Second, previous CN observations have typically targeted the $N=2-1$ lines. Specifically, both~\citet{guilloteau13} and~\citet{reboussin15} quote integrated fluxes for the $N=2-1$, $J=5/2-3/2$ components, which need to be related to the transition observed here. The CN $(N=2-1,~J=5/2-3/2)~/~(N=3-2,~J=7/2-5/2)$-ratio is partially dependent on the disk (temperature) structure. Based on the grid of models discussed in \citet{cazzoletti18}, this ratio is found to vary between 1.1 and 2.1, depending on the disk flaring angle and scale height, but only weakly on mass. For IM Lup, a detailed model of the disk structure exists, based on both gas and dust observations \citep{cleeves16}. For this disk, with flaring angle $\psi=0.15$ and scale height $h_c=0.12$, we can infer a scaling ratio of 1.4, and an $N=3-2,~J=7/2-5/2$ flux of $1.4 \pm 0.1$ Jy\,km\,s$^{-1}$.
	
	Since such detailed models are not generally available, a single scaling factor was used here to obtain estimated CN $N=3-2,~J=7/2-5/2$ fluxes from these surveys, assuming an intermediate value of 1.6. Where appropriate, the typical spread of values introduced by this assumption is indicated in the Figures by arrows to either side. It is important to note that -- while some transition disks are included -- these were not included in the models used to derive the flux ratios used here. However, transition disks with large inner cavities are rare enough \citep[making up $\sim 11 \%$ of the Lupus disks][]{vandermarel18} that we do not expect this to significantly affect our results.
			
	The results of three different CN surveys are shown in Figure~\ref{fig:cndistro}. Using a logrank test (applicable to unbiased data with upper limits), we cannot identify a significant difference in the CN flux distribution between the disks in Lupus (this survey) and Taurus-Auriga \citep{guilloteau13} ($p = 0.064$) if we use $p = 0.05$ as the rejection criterium, and the difference becomes significant only if we take an $(N=2-1,~J=5/2-3/2)~/~(N=3-2,~J=7/2-5/2$) flux ratio of 1.1. Taurus-Auriga seems to be relatively richer in intermediate-luminosity disks, which drive most of this trend. It is however possible that the disks sampled in the \citet{guilloteau13} survey are biased towards those disks in which both continuum and lines are more readily detected, leading to apparently brighter disks in CN. This seems to be the case for the $\rho$ Ophiuchus sample; comparing it to the resolved ALMA images in~\citet{cieza18} reveals a trend towards more radially extended disks, with only two objects included in the CN survey being unresolved. In Taurus-Auriga, such a bias would lead to the difference with Lupus becoming less significant, especially at lower CN luminosities. Comparing the Lupus CN results to those of the $\rho$ Oph disk sample presented in \citet{reboussin15}, however, the difference in flux distributions is always significant: $\rho$ Oph has fainter overall CN emission, and a lower detection rate. A bias towards more radially extended disks in this region would make the lack of CN emission more significant, since we would expect such disks to be brighter rather than fainter on average (see Section~\ref{sec:cnvsr}, below).
	
	\subsection{CN models in the context of the Lupus population}
	\label{sec:cnvspop}
	The Lupus disk survey is uniquely suited to testing models of CN production in protoplanetary disks, both by looking at individual resolved objects and by comparing the full sample properties to the results from a large grid of models. In this Section, this model grid is generated by the \texttt{DALI} physical-chemical disk code, which produces chemical abundances, performs non-LTE excitation calculations, and produces ray-traced images. As mentioned, \citet{cazzoletti18} find several key results for CN emission in disks in their models made with this version of the code: first, an increasing integrated CN flux with increasing disk size; second, a strong dependence of flux on the UV radiation field of stars; and a relatively weak dependence on total disk mass or volatile carbon abundance.
	
	Here, a small~\texttt{DALI} grid is used for comparison with our data, and the results of two previous surveys by \citet{guilloteau13} and \citet{reboussin15}. Since both the Lupus data and the previous surveys give fine-structure fluxes, we used CN the collisional rates from~\citet{lique10}, and raytrace the $N=3-2,~J=7/2-5/2$ and $N=2-1,~J=5/2-3/2$ lines. The grid is constructed using the standard viscous disk model of \citet{lyndenbell74} as used, for example, in \citet{andrews11}, and samples disks with critical radii $R_c$ in the gas disk between 15 -- 60\,AU, and between $2\times10^{-6} - 9\times10^{-2} \,M_{\odot}$ in total disk mass. The stellar spectrum is given by a 4000\,K black-body spectrum emitted by a $1.65\,R_{\odot}$ star; the grid includes models with UV excess (spanning two orders of magnitude in flux) as well as models without excess UV added to this stellar spectrum. The disk structure does not vary in these models: the surface density parameter $\gamma = 1$, while the scale height at 60\,AU $h_c = 0.1$ and the flaring angle $\psi = 0.2$. While the CN flux is sensitive to these parameters, their effect is smaller than that of the varying UV luminosity used in these models \citep{cazzoletti18}, especially since the variation in flaring angles seems to be fairly small \citep{bustamante15}. Although our disks are not in hydrostatic equilibrium, the range of scale heights used here generally reproduces the observed SEDs of disks~\citep{nienke16b}.
	
	In order to compare the models to the observational data, several assumptions must be made. First, a disk (gas) mass tracer is needed. Since $^{13}$CO is not be a good tracer of the total gas mass, we will assume that the gas-to-dust ratio of these disks is $100$ throughout, and that the dust is optically thin, leading to the same approximation used in \citet{ansdell16}:
	
	\begin{equation}
	\label{eq:massdisk}
	M_{disk} = 100 \frac{F_{890\,\mu\rm{m}} d^2}{\kappa_{\nu} B_{\nu}(T_{\rm{dust}})}
	\end{equation}
	
where the factor 100 comes from the gas-to-dust ratio (assumed to be constant and equal to the ISM value), $F_{890\,\mu\rm{m}}$ is the $890\,\mu$m flux, and $d$ is the distance (typically around 160\,pc, and taken from the {\it Gaia} DR2 results for the individual objects in the sample). $\kappa_{\nu}$ is the grain opacity (assumed to be $10$\,cm$^{2}$\,g$^{-1}$ at 1000 GHz with a power-law index of $\beta=1$ \citep{beckwith90}). The $T_{\rm{dust}}$ used here is 20\,K.
	
	The assumed initial abundances of carbon and oxygen are also potentially important, in the context of volatile carbon depletion found in protoplanetary disks \citep[e.g.][]{bergin13,favre13,mcclure16}. However, for CN this is less relevant. For all disk models discussed here, the initial overall abundances are the same as those used in the default models of \citet{cazzoletti18}, since the CN abundance was found not to be sensitive to the depletion of volatile C and O.
	
	A final important assumption is the strength of the UV radiation field. As \citet{cazzoletti18} discuss, CN is very sensitive to the local abundance of excited molecular hydrogen $\rm{H}_2^{*}$, which in turn is sensitive to FUV line pumping. However, typically, no UV spectra exist for these sources \citep[excepting RU Lup and RY Lup][]{ardila13, france14, arulanantham18}. We use a range of UV luminosities, where we focus on the region between $91.2 - 110$\,nm. In that wavelength range, the UV luminosities sampled are between $1.5\times10^{-4} - 1.5\times10^{-6}\,L_{\odot}$ for models with a UV excess, versus $2.3\times10^{-11}\,L_{\odot}$ for models with a purely stellar radiation field. Using a rough approximation, this UV field can be linked to the accretion rate onto the star. Following \citet{kama16b}, the UV is produced from the accretion rate, the stellar mass, and a 10000\,K blackbody spectrum for the accreting material. Using this formulation an accretion rate of $10^{-8}\,M_{\odot}$\,yr$^{-1}$ corresponds to a UV luminosity of $1.5\times10^{-6}\,L_{\odot}$, but as will be shown below, this may be uncertain by as much as an order of magnitude.
	
	\subsubsection{Linking CN to disk radii}
	\label{sec:cnvsr}
\begin{figure*}[t]
	\begin{center}
		\includegraphics[width=\textwidth]{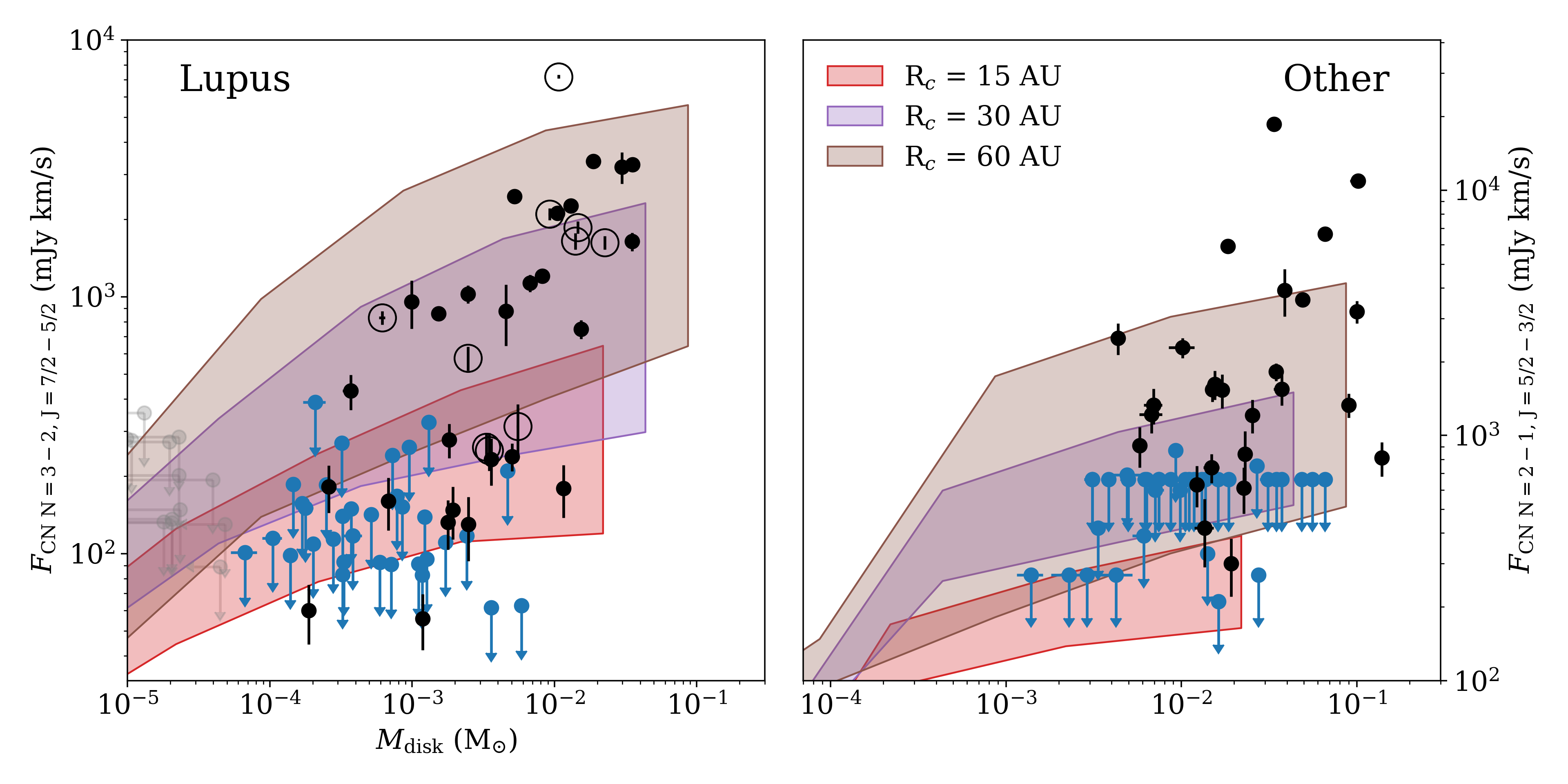}
		\caption{Disk-integrated CN fluxes versus disk masses for the Lupus disk sample {\it(left panel)} and the previously observed disks in Taurus-Auriga and $\rho$ Oph \citep{guilloteau13,reboussin15} {\it(right panel)}. Detected sources are black, CN flux upper limits are blue, and double upper limits are gray. Colored regions indicate model results for the areas of parameter space covered by different UV fluxes and disk masses for a given $R_{\rm{c}}$. Transition disks are marked with a circle and were not modeled. Fluxes for the Taurus-Auriga and $\rho$ Oph sources and models are scaled to a 140\,pc distance.}
		\label{fig:cnvsmod}
	\end{center}
\end{figure*}

\begin{figure}[ht]
	\begin{center}
		\includegraphics[width=0.5\textwidth]{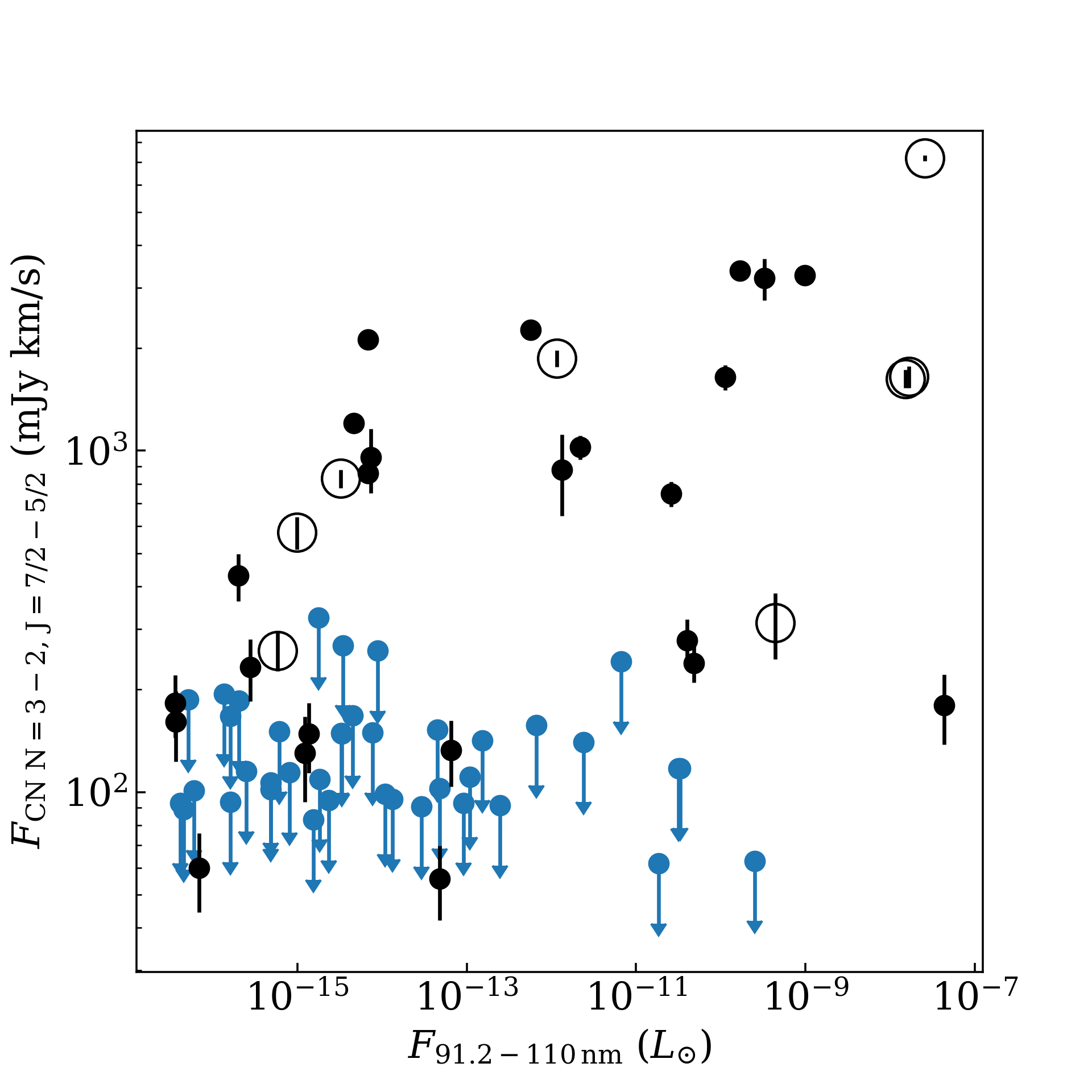}
		\caption{Disk-integrated CN fluxes versus inferred UV fluxes between $91.2 - 110$\,nm from the combined stellar luminosity and the accretion rate onto the star \citep{alcala17}. Circles indicate transition disks.}
		\label{fig:cnvsuv}
	\end{center}
\end{figure}	

	In Figure~\ref{fig:cnvsmod}, the model grid is overplotted to both the Lupus data and the previous large samples presented in \citet{guilloteau13} and \citet{reboussin15}, after raytracing the appropriate transitions. The left panel shows the Lupus data and the corresponding CN $N=3-2$ transition from the models; on the right panel the $N=2-1$ transition fluxes are shown. In this comparison, no assumptions on the line ratios have to be made, since we directly compare to the model fluxes. The colored areas show models with different UV luminosities and disk masses at given $R_{\rm{c}}$. From this Figure, several conclusions can be drawn.
	
	First, the large and unbiased disk population sampled in Lupus makes it apparent that for many of these disks the critical radius $R_{\rm{c}}$ must be small, even less than 15\,AU, in order to reproduce the observed low CN fluxes. This result holds even when considering the impact of large variations in UV luminosity, and when considering variations in disk structure, as in \citet{cazzoletti18}: the faintest detections and non-detections require compact gas disks. In the \texttt{DALI} models, the physical cause behind this strong effect is that, as the disk becomes more compact, the radius at which the CN emission would otherwise peak is located further out than the bulk of the disk material: the disk runs out before the CN can form in significant quantities. 
	
	Meanwhile, in line with \citet{guilloteau98} and the model results in \citet{cazzoletti18}, the brightest disks in CN clearly require large radii, and values of $R_{\rm{c}}$ larger than $60$\,AU, suggesting that the observed Taurus sample contains more massive, but also more radially extended sources than are found, on average, in Lupus.
		
	The observational determination of a disk's $R_{\rm{c}}$ from resolved images is challenging, both for line- and continuum data. Even at high $S/N$, emission below the noise level may be missed. Moreover, gas- and continuum-derived outer radii for the same disk can differ: if a gas line is optically thick it may be detected further out than the dust (by a factor up to 3) \citep{facchini17}. The inward drift and growth of dust grains further change the appearance of the continuum, while leaving the gas unaffected. Both of these effects have been shown in the Lupus star forming region~(\citealt{ansdell18}, Trapman et al. in prep.). While a population of unresolved or marginally resolved dust disks had previously been identified in the uniformly deep continuum survey, our CN models allow us to conclude that their underlying density structure also has a compact {\it gas} surface density distribution when compared to well-studied disks like IM Lup (which has an $R_{\rm{c}}$ of 60\,AU) or TW Hya ($R_{\rm{c}} = 35$\,AU) \citep{cleeves16,huang18}.
	
	\subsubsection{CN as a UV tracer}
	\label{sec:cnvsuv}
	The models run in \citet{cazzoletti18} and the grid used here both predict stronger CN emission from disks around stars that are more luminous in the UV, due to the sensitivity of the CN chemical network to the abundance of $\rm{H}_2^{*}$. With the (known) stellar properties and accretion rates for the disk-bearing stars in Lupus, and the approximation for the UV excess introduced above for the~\texttt{DALI} models, we can calculate a $91.2 - 110$\,nm UV flux for each object in the sample. In Figure~\ref{fig:cnvsuv}, these UV fluxes are compared to the observed CN flux. It is clear that, in contrast to the model predictions, no correlation exists between the CN flux and the UV flux estimated from the stellar flux and accretion rate, and this is confirmed by using a modified Kendall $\tau$-test.
	
	This apparently contradictory result, however, can be reconciled with the model predictions. The accretion rates used to infer the UV flux of these disks are accurate up to a factor of a few, being based on the Balmer continuum excess between $\sim 320 - 346$\,nm and optical line emission \citep{alcala17}. More importantly, while the total excess UV should be proportional to the accretion rate $\dot{M}_{\rm{acc}}$, it is not only emitted as continuum but also as line emission, which is more difficult to constrain \citep[e.g.][]{herczeg04,bergin04}. In particular, the lines between $91.2-110$\,nm are relevant, since they may overlap with UV pumping lines for $\rm{H}_2$ to create $\rm{H}_2^{*}$ via UV pumping. From previous studies of the UV lines of protoplanetary disks at these wavelengths, strong emission lines in this wavelength range have been shown to be present, and responsible for a significant fraction of UV radiation \citep[e.g.][]{herczeg05, ardila13, france14}. 
	
	One way to provide extra, independent information on the UV excess is to use the spatially resolved data, as opposed to the disk-integrated fluxes discussed here. This is because, as demonstrated in Figure 11 of \citet{cazzoletti18}, the radial intensity profile of CN is also very sensitive to the UV radiation field. In the following, two disks for which CN is spatially resolved are discussed in greater detail.

\section{CN images}
\label{sec:twodisksimage}

\subsection{Observations of Sz 98 and Sz 71}
\begin{table}[htb]
\caption{Stellar properties of Sz 98 and Sz 71}
\label{tab:1srcprp}
	
	\centering
	
	\begin{tabular}{l l l l l l}
	\hline \hline
		& SpT & $d$ & $L_\star$ & $M_\star$ & $\log\left(\dot{M}_{\rm{acc}}\right)$ \\
		&    & [pc] & [$L_\odot$] & [$M_\odot$] &  $M_\odot\,\rm{yr}^{-1}$\\
		&	 & (1) & (2) & (2) & (2) \\
	\hline
	Sz 71 & M1.5 & 155.89 & 0.33 & 0.42 & -9.03 \\ 
	Sz 98 & K7 & 156.22 & 1.53 & 0.74 & -7.59 \\
	\hline
	\end{tabular}
	\tablebib{(1) \citet{sfr_book}; (2) \citet{alcala17}}
\end{table}

\begin{table*}[htb]
\caption{Disk emission properties for Sz 71 and Sz 98.}
\label{tab:2fluxes}

	\centering
	
	\begin{tabular}{l l l l l l l}
	\hline \hline
		& $F_{345~\rm{GHz}}$ & $F_{^{12}\rm{CO}~2-1}$ & $F_{^{13}\rm{CO}~3-2}$ & $F_{\rm{CN}~3-2}$ & PA\tablefootmark{a} & $i$\tablefootmark{a} \\
		& [mJy] & [Jy km s$^{-1}$] & [Jy km s$^{-1}$] & [Jy km s$^{-1}$] & [$^{\circ}$] & [$^{\circ}$] \\
	\hline
	Sz 71 & $166.04 \pm 0.63$ & $2.86 \pm 0.26$\tablefootmark{b} & $1.3 \pm 0.1$ & $2.35 \pm 0.95$ & -37.5 & 47.0 \\
	Sz 98 & $237.29 \pm 1.42$ & $2.99 \pm 0.22$\tablefootmark{b} & $0.51 \pm 0.1$ & $3.49 \pm 0.13$ & 107.4 & -40.8 \\
	\hline
	\end{tabular}
	\tablefoot{\tablefoottext{a}{Derived from the continuum data using the \texttt{CASA} \textit{imfit}-task }~\tablefoottext{b}{Affected by foreground absorption and thus a lower limit.}}

\end{table*}

\begin{figure*}[ht]
\begin{center}
		\includegraphics[width=17cm]{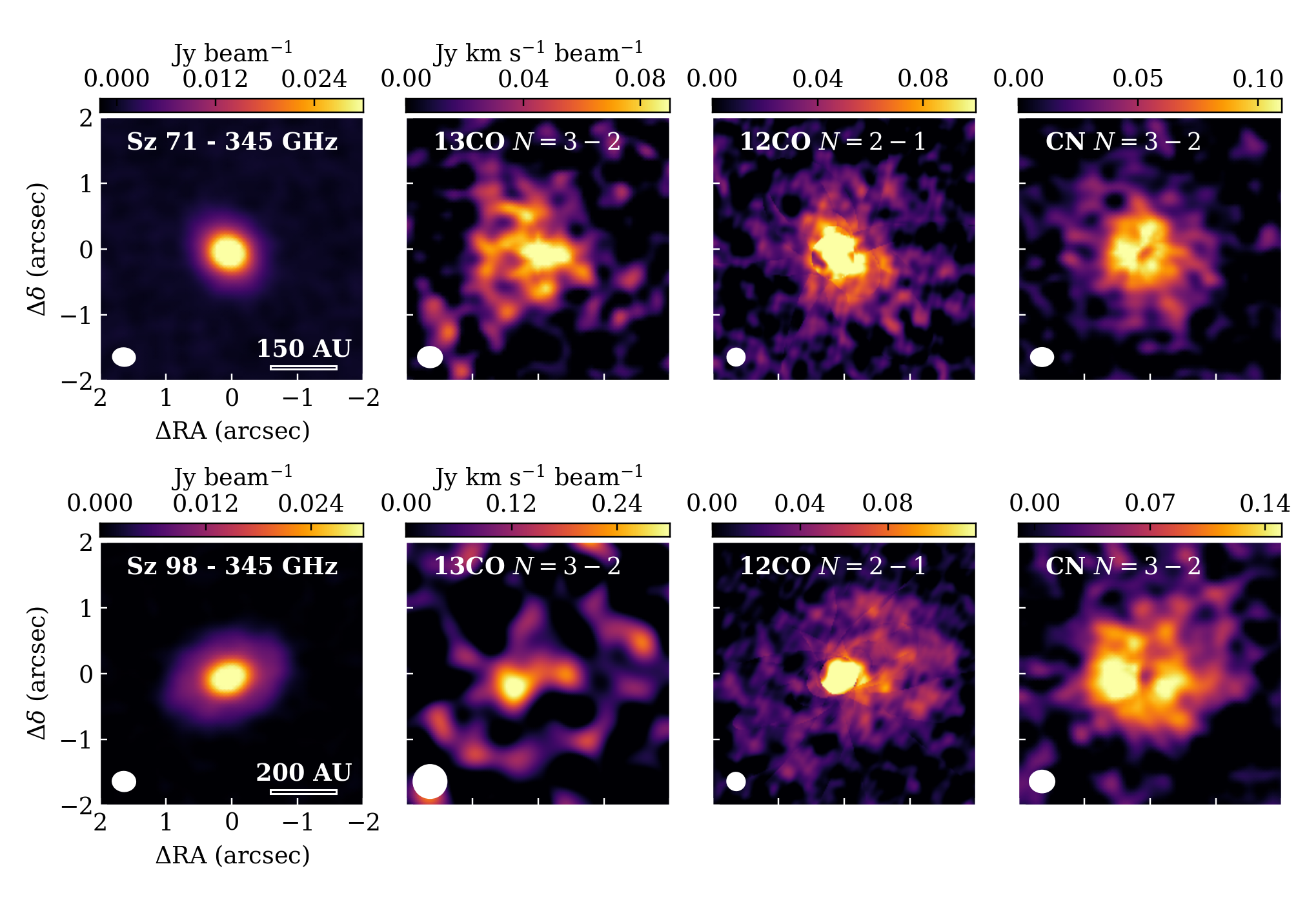}
		\caption{Continuum and moment-zero maps of $^{13}$CO 3-2, $^{12}$CO 2-1, and CN $N=3-2$ (including only the brightest $J=7/2-5/2$ transitions) for both sources.}
		\label{fig:1a}
\end{center}
\end{figure*}

\begin{figure*}[ht]
\begin{center}
		\includegraphics[width=17cm]{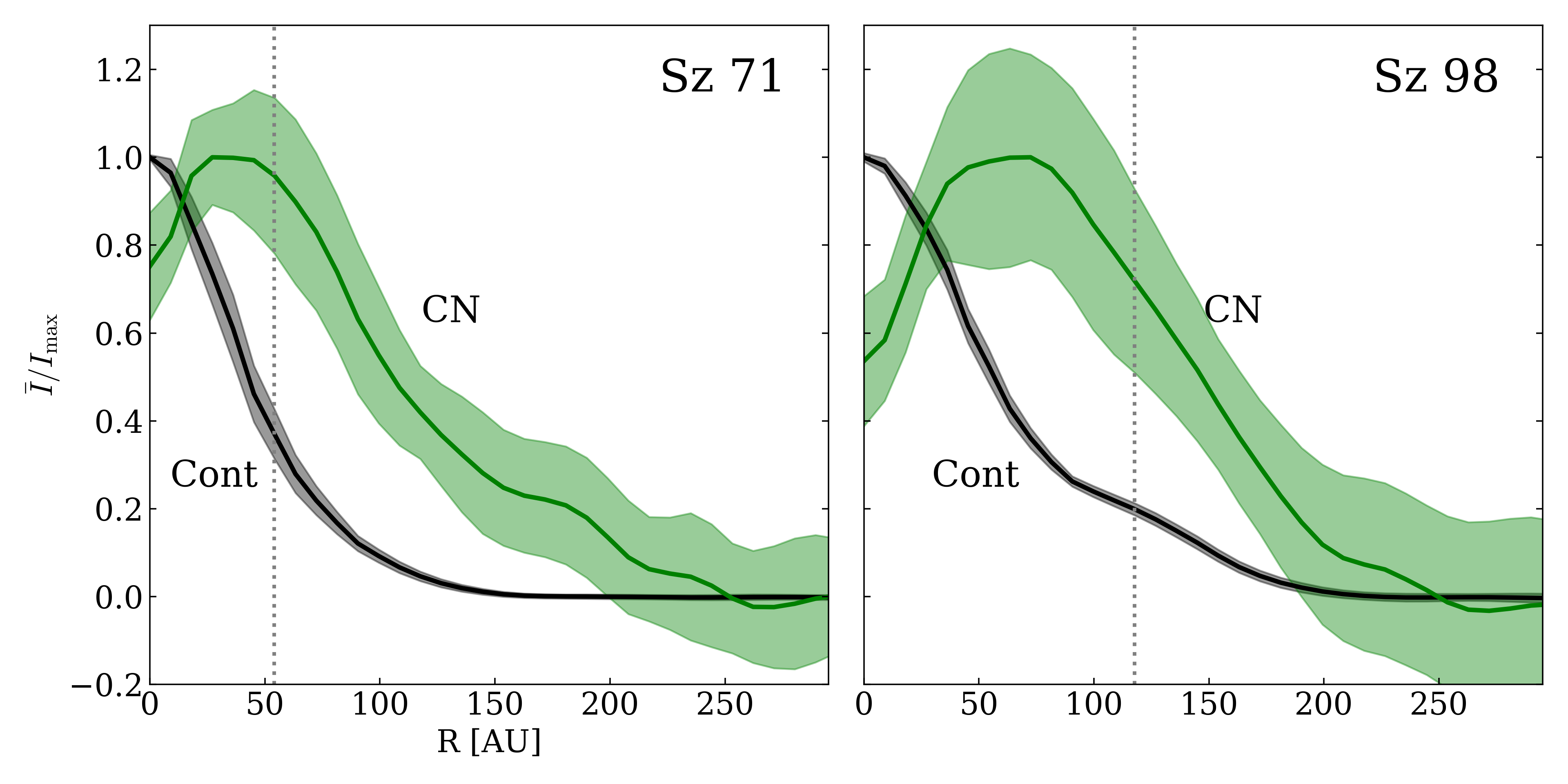}
		\caption{Radial cuts of the continuum (black) and CN moment-zero (green) maps for both sources, normalised to the peak radially-averaged intensity, after deprojecting and derotating the image-plane data. The shaded regions indicate the $1\sigma$ noise level. The gray dotted line indicates the location of a local maximum in the continuum due to an (unresolved) dust ring in Sz 98, and the location of continuum knee in Sz 71 (see Appendix~\ref{sec:uvanalysis}).}
		\label{fig:1b}
\end{center}
\end{figure*}

	The targets selected for detailed discussion in this Section, Sz 98 and Sz 71, are the second and sixth brightest object in CN, respectively. These disks are particularly suited to modeling: their data have high $S/N$ and show clear evidence of a ring-like emission morphology, as well as well-resolved emission. Based on their continuum emission, both disks have a favorable, relatively face-on inclination ($45^{\circ}$ for Sz 98, $36.5^{\circ}$ for Sz 71). The stars do not cover the peak of the spectral type distribution in Lupus, which lies between M3 and M4: Sz 71 is an M1.5 star and Sz 98 is a K7 star.
	
	Figure~\ref{fig:1a} presents the moment-zero maps as well as the disk continuum, showing CN and the bulk-gas tracers $^{13}$CO $J=3-2$ and $^{12}$CO $J=2-1$. In Figure~\ref{fig:1b}, the radial intensity profiles of the continuum and CN line for both sources are shown. Both sources show CN rings, while Sz 98 shows evidence of significant, but non-coincident, continuum substructure. To improve the sensitivity of our analysis, these profiles have been made by azimuthally averaging over the (deprojected) moment-zero maps. In Figure~\ref{fig:2} the spectra of these lines are shown; channel maps of the CN emission towards both sources can be found in Appendix~\ref{app:chanmaps}. C$^{18}$O is not detected at the $3\sigma$-level towards either source. Table~\ref{tab:2fluxes} shows the integrated fluxes for the lines discussed here, as well as the source Position Angles (PAs) and inclinations derived with the {\it imfit} task in \texttt{CASA}.

	The other CN-bright sources are less suited for this purpose. A large fraction of the sources are too compact for structure in the CN emission to be apparent; $28\%$ of the CN-detected sources are unresolved. Of the rest, many have low signal-to-noise levels that make their detailed image-plane analysis impossible. Some of the brightest disks in CN are transition disks, which present a unique continuum structure with a large, central cavity and are exposed to a different UV radiation field as a consequence. This may change the typical behavior of CN. Of the full disks, V1094 Sco is both radially extended and bright, both in continuum and gas emission, but its CN line is partly resolved-out at the largest scales in low-velocity channels, and optically thick continuum emission in the inner disk may obscure the CN emission in those parts \citep{svt18}. For IM Lup, only Submillimeter Array (SMA) data of CN are available, at much lower resolution, and Sz 133 is edge-on.
	
	\subsubsection{CN 3-2}
	CN 3-2 emission in both disks is ring-like, despite the relatively high noise levels of the final moment-zero maps. The peak emission is found at 40 and 60\,AU for Sz 71 and Sz 98 respectively, although the emission does not completely vanish in the central region of the disk. The CN emission extends to large radii in both sources, and is clearly more extended than the continuum: in Sz 71, CN is detected out to approximately 200 AU, and in Sz 98 it can be seen out to 250 AU, while the continuum outer radii for both disks are 150 AU and 200 AU, respectively. The peak emission radius is further from the star in Sz 98 than in Sz 71. The latter source has the smallest contrast between emission at the disk center and peak intensity of these two sources.
	
	In the moment-zero maps of CN, some asymmetry in the emission can be observed: this is likely due to the second-brightest hyperfine structure component ($J=7/2-5/2$, $F=5/2-3/2$) at 0.5\,km\,s$^{-1}$ from the targeted transition which is blended with the strongest hyperfine transition lines of $F=9/2-7/2$ and $F=7/2-5/2$, contaminating the emission redward of the primary hyperfine line. The theoretical relative intensity is expected to be up to $30\%$ of the primary \citep{kastner14,hilyblant17}. However, this does not appear to significantly impact the radial profile observed for the source, apart from some increase in noise in the most affected areas.
	
	Another possible influence on the CN morphology is the existence of optically thick continuum, blanketing the lines. However, we can exclude this possibility: the observed brightness temperatures at the source center (where such an effect would be the most likely) of 6.0\,K and 4.4\,K for Sz 71 and Sz 98 respectively are too low to be consistent with an optically thick continuum if an effective dust temperature of 20 K is assumed. This midplane temperature value is not unlikely given the relatively large beam size of our observations. Moreover, we do detect CO inside the CN rings in both sources, and unambiguously detect $^{13}$CO in the CN hole location in Sz 71.
	
	\subsubsection{$^{12}$CO 2-1}
	$^{12}$CO is a bright and optically thick line in the atmospheres of protoplanetary disks, but it is strongly affected by absorption by foreground clouds towards both sources discussed here, at several separate velocity ranges. By inspecting the spectra in Figure~\ref{fig:2}, some cloud absorption appears to be present in the $^{12}$CO line. This is consistent with single-dish observations of CO in the Lupus clouds \citep{vankempen07}.	Cloud absorption should not have a similar effect on our CN data: the $N= 3-2$ transition has a much higher critical density (of $n_{\rm{crit}} \sim 6.0 \times 10^6$ cm$^{-3}$) than the low-density foreground cloud, and no evidence of cloud contamination of CN is found by \citet{guilloteau13} in Taurus.
	
	Despite the absorption, there is sufficient signal to infer an outer radius for the $^{12}$CO emission: the low-velocity channels blueward of the source velocity are not very absorbed. The outer radius at which CO is detected is larger than that of the CN emission radius towards both sources. Both Sz 98 and Sz 71 have detectable $^{12}$CO out to about 250 AU, where the outer radius is limited by the $S/N$ of our observations. Thus the gas extends at least a factor of $\sim 2 - 3$ further than the continuum in both disks, which is typical for most Lupus disks \citep{ansdell18}.
	
	\subsubsection{$^{13}$CO 3-2}
	The $^{13}$CO emission towards both sources is surprisingly faint and thus the maps suffer from low $S/N$ \citep{ansdell16, miotello17}. Towards Sz 98, the detection of the line is marginal. Binning the data to 1\,km\,s$^{-1}$-wide channels and lowering the resolution to $0.6''$ by removing short baselines improves our sensitivity. The moment-zero map of $^{13}$CO in this source does not show emission across the full disk, but only in `lobes' towards the sides; this is probably caused by low $S/N$ of the data, as well as the limb-brightening effect of optically thin emission. In Sz 71, the emission is brighter and clearly centrally peaked, but also faint. Because of the low surface brightness of the line and the large $S/N$ difference between the sources, we have not included $^{13}$CO radial cuts in Figure~\ref{fig:1b}.
	
	The faintness of $^{13}$CO in Sz 98 is especially surprising: it is one of the largest and brightest (in both $^{12}$CO and continuum) disks in the Lupus sample, even including transition disks. Comparing the regions of the spectrum where $^{12}$CO is absorbed to the $^{13}$CO spectrum in Sz 71 (Figure~\ref{fig:2}, top) shows that foreground absorption is unlikely to be the full explanation for the faintness of the $^{13}$CO line.

	\subsubsection{CN versus continuum}
	It is possible that -- similar to the situation for CO \citep[e.g.][]{isella16, fedele17} -- a link exists between continuum structure and the radial distribution of CN. Of the two disks discussed here, Sz 98 has the most obvious continuum feature: as shown in Figure~\ref{fig:1b}, based on the $u,v$-plane fit in Appendix~\ref{sec:uvanalysis}, a secondary maximum exists at the location of the `bump' in the continuum at $\sim 120$\,AU. No counterpart of this feature is seen in the CN emission profile: CN peaks at a radius interior to it, and seems to present a fairly constant downward slope outside 100\,AU.

	For Sz 71, the continuum structure is less obvious: around 50\,AU, the continuum profile begins to drop off less steeply. This is possibly related to an unresolved feature, but this is not clear even from the $u,v$-plane. The dashed line in Figure~\ref{fig:1b} for this object indicates the point where the turnover to a less steep profile occurs; it is fairly close to the CN peak.
	
	Given the differences in the comparison between continuum and CN for these two sources, we therefore conclude that no obvious relation between the two is present in all cases. The clear presence of ring-like CN emission profiles in Sz 71 and Sz 98 is consistent with the predictions for rings of CN emission \citep{cazzoletti18}, and suggests that these features may be the result of chemistry alone, without the need for underlying continuum structures.

\subsection{\texttt{DALI} modeling of CN rings}
	\label{sec:dalianalysis}	
	Here, a purely chemical cause for CN rings is investigated by more detailed modeling of the two bright, resolved disks discussed previously. The thermo-chemical disk modeling code \texttt{DALI} was used to test if simple models without the dust substructures identified in the previous subsection can produce radial distributions and integrated intensities of CN similar to those observed.
	
	 Based on the results of \citet{cazzoletti18}, the presence of CN rings appears to be a common feature for their grid of protoplanetary disk models. Our goal is not to provide a perfect fit to the data, but rather to investigate whether rings similar (in terms of total flux and radial profile) to those observed in these two disks can be produced without continuum substructure, and to isolate the main parameters governing CN emission strength and morphology. It is important, in this context, to note that in an absolute sense the uncertainties resulting from the assumptions in the chemical network alone, as well as from disk structure, lead to an uncertainty of a factor of a few in flux \citep{woitke18}.
	
	\subsubsection{Individual disk model details}
	In order to find a rough physical structure of the disk models, we first attempted to reproduce the objects' SED, described in more detail in Appendix~\ref{app:sedfit}. This ensures that the large-scale distribution of the dust temperatures in the \texttt{DALI} models is similar to that of the source, and allows us to find, for instance, the CO snowline in the disks. Again, a parametric dust distribution model, as in \citet{andrews11}, was used, constrained by data spanning a wide range of wavelengths between visible and mm-wavelengths. The gas surface density is $100 \times$ the dust surface density and is not decoupled. The disk structure parameters used for the models of both disks have not been varied further in their respective \texttt{DALI} models and can be found in Table~\ref{tab:appmods}. The outer radii of the disks were taken to be $200$\,AU, based on the CN data. The stellar parameters adopted for the input photosphere in the models were taken from \citet{alcala14,alcala17}. As before, an initial estimate of the relevant UV flux for CN formation, $L_{912-1100\, \textrm{\AA}}$, the UV excess emission from the model star was parametrized by the mass accretion rate, assuming $10000$\,K blackbody emission from the accreted material; due to the aforementioned uncertainties we have allowed for variations in the UV excess at these wavelengths of up to a factor 10.
	
	A model of Sz 71 with a UV-excess of $\log(L_{912-1100\, \rm{\AA}} / L_{\odot}) = -5.2$, based on the measured accretion rate, as well as a model with an increased UV excess luminosity of $\log(L_{912-1100\, \rm{\AA}} / L_{\odot}) = -4.1$, denoted as M1 and M2 respectively, are presented in this Section. For Sz 98, we only ran models with a single UV luminosity based on the observed accretion properties, and thus only have an M1, since this disk model already performed quite well in reproducing the CN flux and radial profile. The surface density profiles were kept fixed for all models of a single source. After ray-tracing each model, the images were convolved with an $0.3''$ circular Gaussian beam, to see if the CN emission was similar to the observations.
	
	\subsubsection{Model results}
\begin{figure*}[ht]
\begin{center}
	\includegraphics[width=17cm]{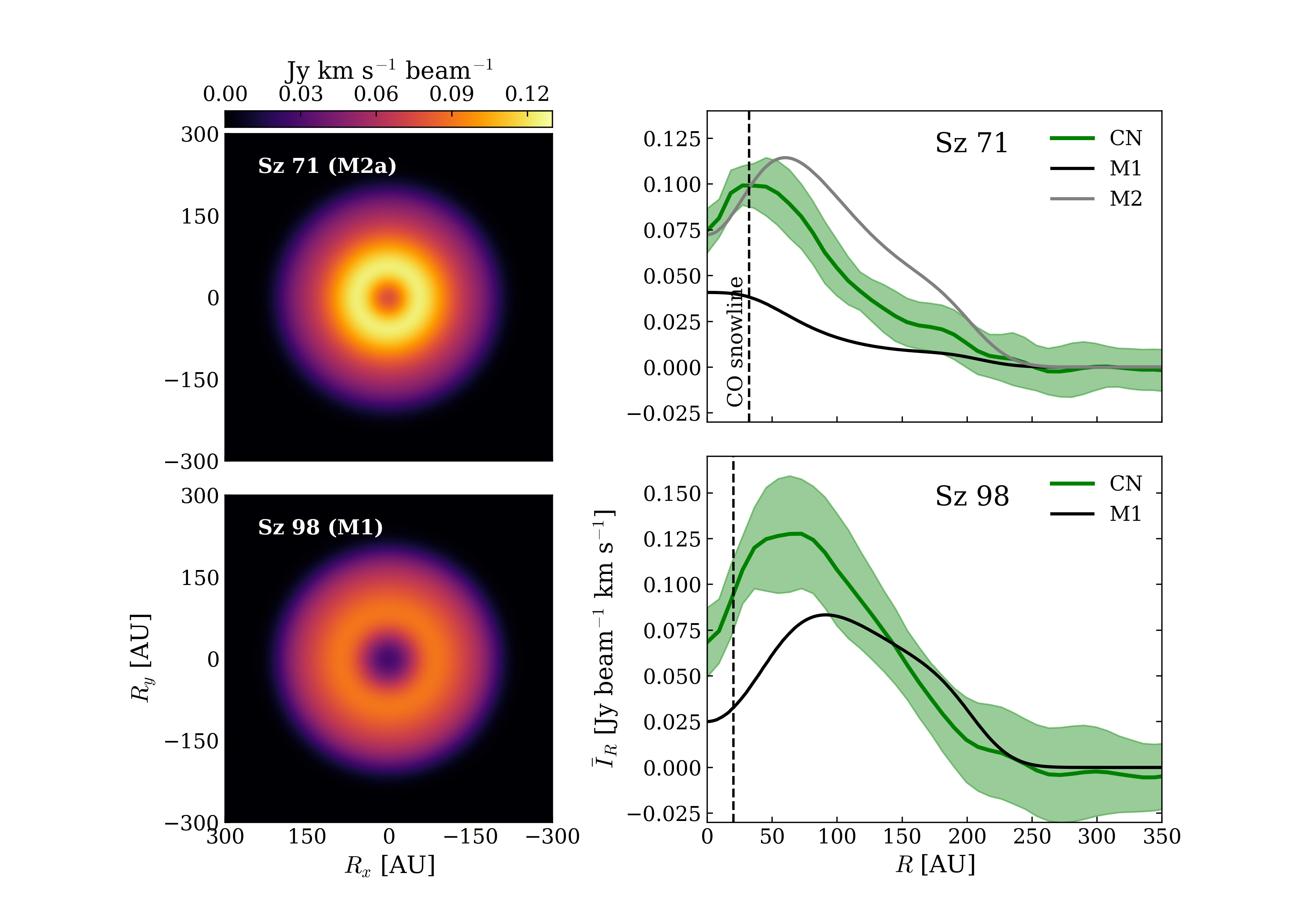}
	\caption{CN model images and radial intensity profiles (black and gray) from the \texttt{DALI} models, after convolving with a $0.''3$ beam for both sources. Model M2 for Sz 71 has an increased UV luminosity relative to M1. The black dashed vertical lines indicate (approximately) the CO snowline locations in the models. The observed radial profiles of CN are shown in green, together with their $1\sigma$ errors.}
	\label{fig:4}
\end{center}
\end{figure*}

	As can be seen in Figure~\ref{fig:4}, the most remarkable feature of the CN emission, its concentration in a ring, is indeed reproduced in the model radial profiles, and occurs at radii within 20 AU of those seen in the data for Sz 71 model M2 and Sz 98 model M1. Also, the peak intensity of the emission is close (within $40\%$) to the observed peak value for the most similar models. The similarity in the emission profiles outside the peak is also encouraging, especially considering the factor few uncertainty in absolute fluxes from this type of chemical model. 
	
	Table~\ref{tab:4cnmods} presents the total CN fluxes for the different models. The agreement is closest for Sz 98: the model flux and observed flux only differ by $5\%$. For Sz 71, if an increased UV luminosity is used, a larger difference can be seen, of $\sim 60\%$ for all models; but the emission profile for M2 is clearly closer than that of model M1 for this disk, primarily due to the increase in peak intensity radius with increased UV flux.
	
	No clear link appears between the CO snowline position \citep[defined as the location where the disk midplane reaches 20\,K, following][]{oeberg05} and the location of the CN peak in Sz 98. For Sz 71, the CO snowline, CN peak, and -- possibly -- the continuum slope change lie within 10 AU, but the location of the snowline is quite uncertain, and both it and the continuum structure may not be closely related to the CN emission, which generally arises from the upper parts of the disk in models \citep{cazzoletti18}. In either case, the significant difference between the radii of the expected CO snowline position, the CN peak, and the continuum structure in Sz 98 seem to preclude a clear, one-to-one link between these quantities in all disks.

\begin{table}[hbp]
\caption{Parameter values and integrated CN fluxes for the \texttt{DALI} models of Sz 71 and Sz 98; the most representative model is indicated in \textbf{boldface}.}
\label{tab:4cnmods}
	\centering
	\begin{tabular}{l l l l}
	\hline \hline
		  & 	& $\log(L_{912-1100\,\rm{\AA}} / L_\odot)$  & $F_{\rm{CN}}$	\\
		  &     			&       & [Jy\,km\,s$^{-1}$] \\ \hline
	Sz 71 & M1				& -5.2 	& 0.8 \\
		  & \textbf{M2}		& -4.1 & 3.9 \\
 		  & \textit{Obs}	& 	   & 2.4 \\
 		  
	Sz 98 & \textbf{M1}		& -3.4 & 3.7 \\
		  & \textit{Obs}	&      & 3.5 \\
	\hline
	\end{tabular}
	
\end{table}

	\subsubsection{CN rings as a disk probe}
	\label{sec:cnvscont}
	The main factor determining the CN radius and emission strength in these large-disk models is the excess UV flux due to accretion of material onto the star. This is especially obvious from comparing models M1 and M2 for Sz 71. It is crucial to note that in none of our models we need the presence of substructure in the disks' surface density, either in the dust or gas components: CN rings arise purely from a chemical effect. 
		
	In the wider grid of disks in \citet{cazzoletti18}, as well as in these models, CN rings are identified as a common feature. However, the ring radii and fluxes decrease with stellar $T_{\rm{eff}}$ and disk size, which explains why we can only study them in these two sources in the survey: below $R_c = 40$\,AU, which applies to most of the objects in our sample, the ring radius drops rapidly and starts to become unresolved at our observations' effective resolution. At the same time, the flux -- and therefore $S/N$ of the resolved CN images -- fall off. In the grid of models discussed in Section~\ref{fig:cnvsmod}, the smallest CN peak radii would be resolved with an effective beam size of $\sim 10$\,AU, given a face-on inclination and sufficiently good $S/N$.  
	
	Ultimately, CN is best used as a probe of the $91.2 - 110$\,nm UV radiation field impacting the upper disk. This is especially the case if the other disk parameters, particularly the characteristic radius $R_c$, can be constrained independently. For this, other common molecular tracers like $^{12}$CO or $^{13}$CO can be used, leaving the way open for such studies in the future.

\section{Conclusions}
	Our ALMA observations of a $99\%$ complete sample of Class II disks in Lupus allow us to study the behavior of CN, both in the full sample and -- using individual models -- in two resolved sources, and to use this common molecule as a probe of upper-disk atmosphere properties.	With higher resolution and deeper observations, this molecule can be used to gain essential information on the properties of protoplanetary disks. The main results obtained from the observations discussed here are as follows:
	\begin{itemize}
	\item The CN $N=3-2$, $J=7/2-5/2$ transition is bright, and has a similar detection rate to $^{13}$CO in disks in Lupus, but is not strongly correlated with either $^{13}$CO or continuum fluxes in these disks.

	\item Comparing the CN flux distribution in Lupus to that of a population of Taurus-Auriga disks shows no significant difference, while disks in $\rho$ Oph may be fainter on average.

	\item Comparing the CN fluxes of the full Lupus sample to a model grid with varying UV fluxes and gas disk characteristic radii $R_c$ supports the conclusion that a significant number of disks in Lupus has a compact gas surface density profile, with $R_{\rm{c}} \leq 15$\,AU needed to explain part of the population.

	\item CN shows a ring-shaped emission morphology towards the bright, resolved sources Sz 71 and Sz 98, consistent with model predictions that CN is generally distributed in rings. For the bulk of objects, it is not possible to detect rings at this sensitivity and resolution, consistent with the expectations for compact, low-mass disks around late-type stars.

	\item Continuum substructure unrelated to the CN rings appears to be present in both sources, and particularly in Sz 98, where we infer an unresolved continuum ring at $\sim 120$\,AU. No connection between the CO snowline, the continuum structure, and the CN ring are seen in Sz 98.
	
	\item Disk models without substructure retrieve CN rings similar to those observed towards both sources, in terms of both peak emission radii and absolute intensities. Their radii strongly dependent on the amount of excess UV emission. This implies that CN rings are purely chemical in nature, and good tracers of the effective UV irradiation of the upper disk atmosphere, especially if gas disk sizes are determined from other data.
	\end{itemize}

\begin{acknowledgements}
We kindly thank J. M. {Alcal{\'a}} for providing us with X-shooter spectra of both sources, and A. Bosman, and D. Harsono for useful discussions. We also thank S. Bruderer for his role in the continued development of \texttt{DALI} and his advice on how best to use it. Astrochemistry in Leiden is supported by the European Union A-ERC grant 291141 CHEMPLAN, by the Netherlands Research School for Astronomy (NOVA), and by a Royal Netherlands Academy of Arts and Sciences (KNAW) professor prize. This paper makes use of the following ALMA data: ADS/JAO.ALMA\#2013.1.00220.S and ADS/JAO.ALMA\#2015.1.00222.S. ALMA is a partnership of ESO (representing its member states) NSF (USA) and NINS (Japan) together with NRC (Canada) MOST and ASIAA (Taiwan), and KASI (Republic of Korea), in cooperation with the Republic of Chile. The Joint ALMA Observatory is operated by ESO, AUI/NRAO and NAOJ. The authors acknowledge support by Allegro, the European ALMA Regional Center node in The Netherlands, and expert advice from L.T. Maud in particular. MT has been supported by the DISCSIM project, grant agreement 341137 funded by the European Research Council under ERC-2013-ADG. SF, CFM and AM acknowledge an ESO Fellowship.
\end{acknowledgements}

\bibliographystyle{aa}
\bibliography{LupusCN_def_bib}

\begin{appendix}
\renewcommand\thefigure{\thesection.\arabic{figure}}
\section{CN versus C$^{18}$O in Lupus disks}
\label{app:othertracer}
\setcounter{figure}{0}
\begin{figure}[h]
\begin{center}
		\includegraphics[width=0.5\textwidth]{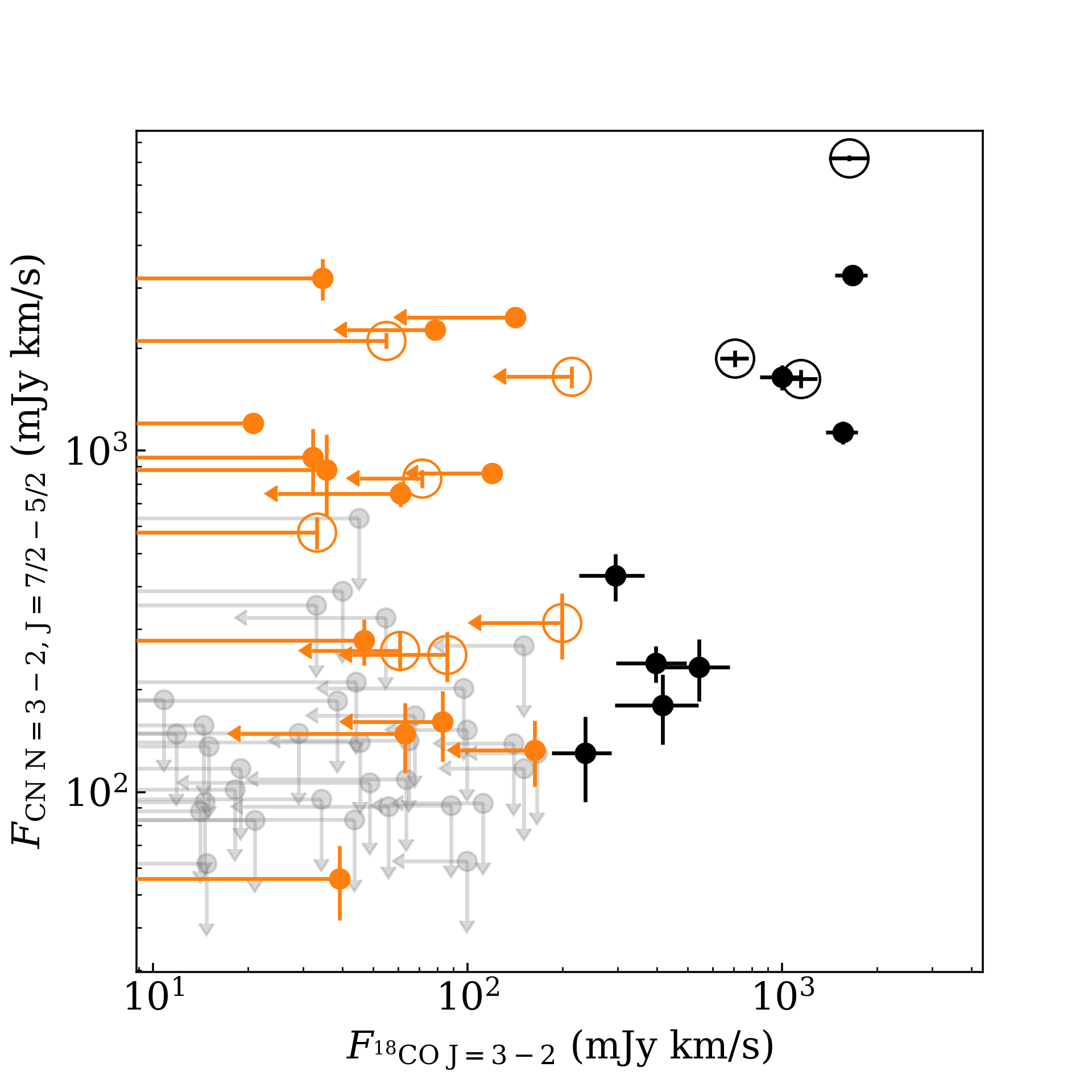}
		\caption{Integrated CN versus C$^{18}$O $3-2$ fluxes for the Lupus disk sample. Detected sources are black,  C$^{18}$O $3-2$ upper limits are orange, CN flux upper limits are blue, and double upper limits are gray. Transition disks are marked with a circle.}
		\label{fig:cnvsc18o}
\end{center}
\end{figure}

\renewcommand\thefigure{\thesection.\arabic{figure}}
\section{$u,v$-plane analysis}
\setcounter{figure}{0}
\label{sec:uvanalysis}
	It is possible that, in the disks in which a clear CN ring is present, some underlying continuum structure is responsible, for instance, by causing a local density enhancement. Therefore, it is relevant to first establish if any such structures are present in the disks we study in detail, and if so, where they are located.
	
	In the two disks discussed here, based on their continuum intensity profile (Fig.~\ref{fig:1b}), Sz 98 seems to have a bright central core, with a large, fainter outer region, giving it something of a `fried-egg'-appearance similar to V1094 Sco \citep{svt18}; the radial profile seems to suggest an unresolved structure in this outer region. Sz 71, on the other hand, may have a `knee' in its intensity profile at around 50\,AU. Analysing the continuum emission in the u,v-plane can provide a more thorough picture of the mm-dust structure underlying the CN emission, even if such features are difficult to detect in the image plane due to the deconvolution algorithm. \citet{tazzari17} performed a fit of both disks with a smooth self-similar disk model, including radiative transfer. However, such a smooth model is not expected to be able to fit radial substructures in the continuum, if any are present. For that reason, a simpler fitting method is used here, focusing only on reproducing the radial intensity profile.
	
	Both sources were deprojected and derotated based on an image-plane fit with the \texttt{CASA} imfit-task. We confirmed that, within the uncertainties given by these values, the results of our analysis did not significantly change. Subsequently the data were binned to $15$ k$\lambda$ bins in u,v-space, and finally scaled and fit with two simple models: a Gaussian core (Model 1), the single-component model, and a Gaussian core modulated with a cosinusoidal term (Model 2), the two-component model, to mimic the behavior of an (unresolved) ring of particles. This method was successfully used by \citet{zhang16} to describe rings in a variety of transitional and full disks, and has the important property that it uses a minimal number of parameters, while being smooth (making it similar to these disks' appearance) and not necessarily containing nulls (which must occur for sharp-edged models).

\begin{equation}
\begin{split}
	I(\theta) = & \frac{1}{\sqrt{2 \pi} \sigma_0} \exp \left( - \frac{\theta^2}{2\sigma_0^2} \right) \\
	& + \sum_j \cos (2\pi\theta\rho_j) \times \frac{a_j}{\sqrt{2\pi}\sigma_j} \exp \left( -\frac{\theta^2}{2\sigma_j^2} \right).
\end{split}
\end{equation}

Here, $a_j$ is the relative intensity of the $j$th component, $\sigma_j$ its width (converted, for ease of reading, to AU in Table~\ref{tab:3uvfit}), and $\rho_j$ the spatial frequency which modulates the $j$th Gaussian, and $\theta$ is the angle to the source center (in radians). Baselines up to 800 k$\lambda$ are used for our fit to Sz 71, and up to 500 k$\lambda$ for Sz 98; beyond these values, noise starts dominating our u,v-plane data. Parameter values for two models, with $j_{\rm{max}}=0$ and $j_{\rm{max}}=1$, were obtained from MCMC-fits to the observations with the~\textit{PyMC} Python module.
	
\begin{figure*}[ht]
\begin{center}
	\includegraphics[width=\textwidth]{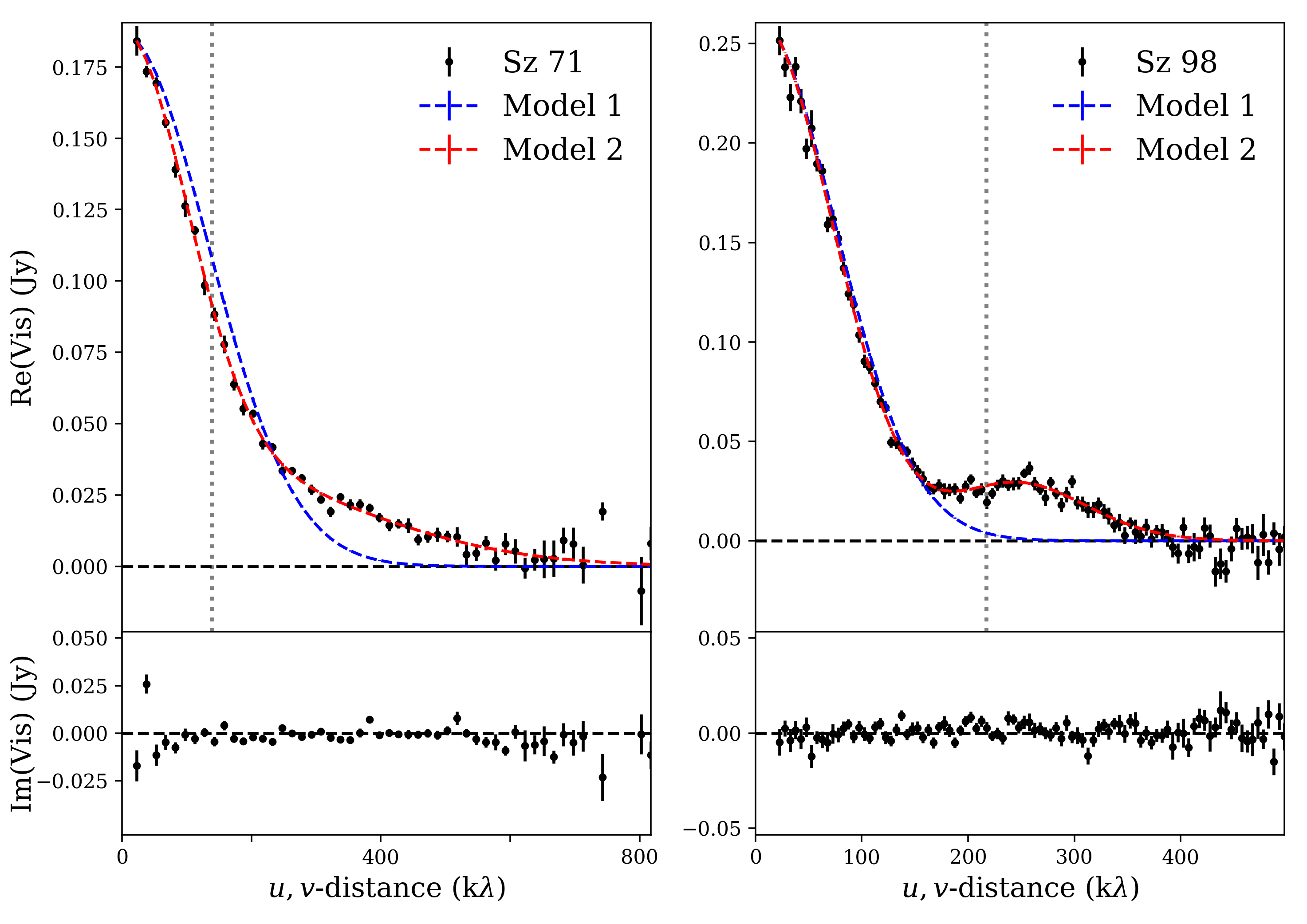}
	\caption{345 GHz continuum visibilities for Sz 71 and Sz 98, after deprojection and derotation. The real visibilities have been overplotted with the best fitted single-component model (Model 1, blue) and the best-fit two-component model (Model 2, red). The best-fit spatial frequency of the two-component model is indicated by the gray dashed lines.}
\label{fig:3}
\end{center}
\end{figure*}

	In Figure~\ref{fig:3} and Table~\ref{tab:3uvfit}, the results of the fitting procedure are shown. For both Sz 71 and Sz 98, the two-component Gaussian model fits the continuum significantly better than the single-component one, when comparing the Bayesian Information Criterion (BIC) \citep{schwarz78}: $\Delta_{\rm{BIC}} = 1296.9$ for Sz 71, and for Sz98, $\Delta_{\rm{BIC}} = 1382.7$ in favor of the two-component model. The two-component model performs especially well at large baselines, suggesting that there is continuum structure at small spatial scales. For Sz 98, a true second maximum is immediately obvious in the visibility plane.The similarity of this feature to those seen (at much higher resolutions and sensitivities) in e.g. \citet{zhang16}, \citet{HLTau} and \citet{TWHya} is remarkable, all the more so considering its amplitude -- not much fainter than the structures identified in the V1094 Sco disk \citep{svt18}. For Sz 71 the best-fit values of $\rho_1$ show a fairly large uncertainty, and the amplitude of the small-scale signal is not very large. 
	
	In Sz 98, we can assign the identified u,v-plane feature to the presence of an (unresolved) ring in the image plane, located at 117\,AU, depicted with a dotted gray line in Figure~\ref{fig:1b} (right); this is consistent with the knee that is tentatively identified in the continuum radial intensity profile. Moreover, \citet{tazzari17} find a clear ring-like residual when comparing their best-fit model of Sz 98 to the data, located in the same part of the disk. The visibility feature in Sz 71 is weaker and overlaps with the bright `core' of the disk; thus, without higher-resolution and deeper data, we can only say that its brightness profile drops off less steeply beyond 50\,AU in this disk (Figure~\ref{fig:1b} (left)). Given the similarity of these disks' visibilities, it is possible that a structure similar to that of Sz 98 is present in Sz 71, but at a fainter level and closer to the central star. However, no ring-like residual is present for this disk in the best-fit smooth disk model by \citet{tazzari17}. It is also important to note that our intensity profiles do not provide any information on the vertical structure of the continuum, since the mm-sized grains responsible for the emission modeled here are expected to be vertically settled with respect to the disk gas \citep{dullemond05,pinte16}.
	
\begin{table*}[hb]
\caption{Parameter values and inferred errors from the MCMC fit for both models and sources.}
\label{tab:3uvfit}

	\centering
	\begin{tabular}{l l l l l l l l}
	\hline \hline
		&	& $\sigma_0$ & $a_1$  & $\sigma_1$ & $\rho_1$ & $\chi^2_{\nu}$ & BIC \\
		&	& [AU] & [Jy beam$^{-1}$] & [AU]         & $[\rm{k}\lambda]$ &  &\\
	\hline
	Sz 71 & Model 1 & $37.1 \pm 0.18$ & - & - & - & 27.87 & 1071.2 \\
		  & Model 2 & $49.8 \pm 0.80$ & $0.969 \pm 0.055$ & $19.9 \pm 1.7$ & $138 \pm 23$ & 1.71 & -225.7 \\
	\hline
	Sz 98 & Model 1 & $87.7 \pm 0.40$ & - & - & - & 20.67 & 1306.8 \\
	      & Model 2 & $85.3 \pm 0.71$ & $0.467 \pm 0.11$ & $91.0 \pm 3.0$ & $217.0 \pm 2.1$ & 1.45 & -75.9 \\
	\hline
	\end{tabular}

\end{table*}

\renewcommand\thefigure{\thesection.\arabic{figure}}
\section{SED fitting procedure}
\setcounter{figure}{0}
\label{app:sedfit}
	We attempted to closely reproduce the SEDs of both sources with our models, in order to find a dust density and temperature structure corresponding as closely as possible to the real (disk-averaged) properties of the disks, especially at mid-infrared down to mm wavelengths. Using the resulting parametrized disks as input for the \texttt{DALI} physical-chemical modeling code allows us to confirm if the CN rings we observe in the sources are the result of just the global properties of the disk (particularly, the interplay between gas density and UV radiation) or if they require underlying dust and/or gas substructures at the radius of the CN ring (neither of which would be revealed by the SED).

	The disk model used is adapted from the model used in \citet{andrews11}, but without an exponential taper at characteristic radius $R_c$. Instead, the disk is truncated at $R_c = R_{\rm{out}}$, and the surface density slope is fixed at $\gamma = 1$ for all models. The parameter space of this model was explored in two phases. For the first phase, we used the \texttt{RADMC3D}-code\footnote{\url{http://www.ita.uni-heidelberg.de/~dullemond/software/radmc-3d/}} to globally constrain the flaring angle of the disk $\psi$, its scale height $h_r$, the disk mass $M_d$, and the possibility of an inner cavity ($R_{\rm{in}} \geq R_{\rm{sub}, \rm{silicate}}$).\par
Taking $R_{\rm{in}}$ as a variable was motivated by our difficulty reproducing the near- and mid-IR parts of the SED when using the standard parametrization of $R_{\rm{sub}, \rm{silicate}}$. However, we only take a values of $R_{\rm{in}}$ in a small interval, as radii beyond $\sim 2$ AU would lead to obviously different SEDs. The shape of the near- and mid-IR SED is also sensitive to the choice of parametrization of the inner region structure: a puffed-up inner rim extending to the sublimation radius, for instance, could lead to a similar SED. However, such changes would make a significant difference only at scales much smaller than our ALMA observations' resolution.

	The values of $R_{\rm{out}}$ and $i$ in our models are taken from the 870 $\mu$m CN and continuum images, respectively. For both disks, we take $R_{\rm{out}}$ to be $\sim 200$ AU. We use a single grain population between 0.5 and 1000 $\mu$m. Our opacities are based on the same abundances of amorphous and crystalline silicate and amorphous carbon grains as in \citet{weingartner01}.
	
	The second phase - performed in \texttt{DALI} - uses the best-fit values from the \texttt{RADMC3D} models, but includes a population of small grains and settled large grains. The fraction of large grains, here, is $f$, and the decreased scale height of the settled large grain population is equal to $\chi h_r$. We explore only $f = [0.85, 0.99]$ and $\chi = [0.2, 1.0]$. Our small grain population is limited to a maximum size of 1 $\mu$m. Including this small grain population means we have a better description of the silicate features in the NIR (especially for Sz 71, for which we have an IRS spectrum). Using \texttt{DALI} also allows us to make sure no gridding inconsistencies are present when we run the full thermo-chemical code to calculate the CN emission.
	
	For the stellar luminosity, effective temperatures, and accretion luminosities we used the values given in Table~\ref{tab:1srcprp}. We corrected the data for extinction using $R_V = 3.1$, and $A_V = 0.5$ mag and $1.0$ mag for Sz 71 and Sz 98 respectively \citep{alcala17}. In the case of Sz 71, the X-shooter spectra have been rescaled by a factor of $1.5$ (within the instrument's systematic error, c.f. \citet{alcala14}) to agree with the 2MASS J, H, K-magnitudes. For Sz 98, we show the 2MASS and B and V-band photometry, but do not include them in our fit. This source is a known variable, and the X-shooter data have the largest amount of wavelength coverage in a single observation. Moreover, the most accurate determination of $T_{\rm{eff}}$ and $L_{\star}$ is based on these data.

	\subsection{SED data}
	The disks' SEDs were constructed on the basis of available literature data. For Sz 98, the dataset consists of data from the X-Shooter (from \citet{alcala17}), WISE, and the Herschel PACS and SPIRE instruments \citep{bustamante15}, as well as 2MASS J, H, K and B and V-band fluxes \citep{2mass}. The PACS data, however, were not used in the fit since it appears to be affected by a systematic offset possibly caused by the photometric calibration.
	
	Sz 71 has not been observed by Herschel, and the behavior of its SED is therefore much less constrained at FIR wavelengths. However, B, V, R-band and X-Shooter~(from \citet{alcala14}) and 2MASS data do exist, as well as a Spitzer IRS spectrum between 10 and 36 $\mu$m \citep{spitzerIRS}. The Sz 71 SED appears to be showing a significant contribution from amorphous carbonaceous grains, in the form of excess emission at wavelengths down to 1 micron. These opacities are not included in our models. However, since their primary effect is to heat the innermost part of the disk (outside our chemical region of interest) we do not expect to seriously underestimate the CN emission by excluding them in our full thermo-chemical models.

\subsection{SED fitting results}
	Both disks appear to be non-flaring compared to e.g. the sample presented in \citet{marel16whole} when fit with a similar methodology and disk model. This is significant, since the flaring angle has a large impact on the dust heating at radii where we expect CN to be present \citep{cazzoletti18}. However, there is still a significant difference between both disks' flaring angles, meaning our results are stil representative of disks in general with a variety of flaring angles.

\begin{table}[ht]
\caption{Parameter ranges and used values for the disk models}
\label{tab:appmods}
	\centering
	%Sz 71:
	\begin{subtable}[b]{0.4\textwidth}
	\begin{tabular}{l c c c}
	Sz 71 & & & \\ \hline \hline
	Parameter & minimum & maximum & used \\
	\hline
	$h_r$ & 0.07 & 0.15 & 0.11 \\
	$\psi$ & 0.1 & 0.3 & 0.14 \\
	$f$ & 0.85 & 0.99 & 0.99 \\
	$\chi$ & 0.2 & 1.0 & 0.2 \\
	$R_{\rm{in}}$ (AU) & 0.04 & 1.5 & 1.0 \\
	$M_d$ ($M_{\odot}$) & 0.01 & 0.09 & 0.03 \\ \hline
	\end{tabular}
	\end{subtable}
	\\
	%Sz98:
	\begin{subtable}[b]{0.4\textwidth}
	\begin{tabular}{l c c c}
	Sz 98 & & & \\ \hline \hline
	Parameter & min & max & best \\
	\hline
	$h_r$ & 0.07 & 0.15 & 0.1 \\
	$\psi$ & 0.07 & 0.1 & 0.085 \\
	$f$ & 0.85 & 0.99 & 0.99 \\
	$\chi$ & 0.2 & 1.0 & 0.2 \\
	$R_{\rm{in}}$ (AU) & 0.06 & 1.5 & 0.5 \\
	$M_d$ ($M_{\odot}$) & 0.01 & 0.09 & 0.07 \\ \hline
	\end{tabular}
	\end{subtable}

\end{table}

\begin{figure*}[ht]
\begin{center}
	\includegraphics[width=\textwidth]{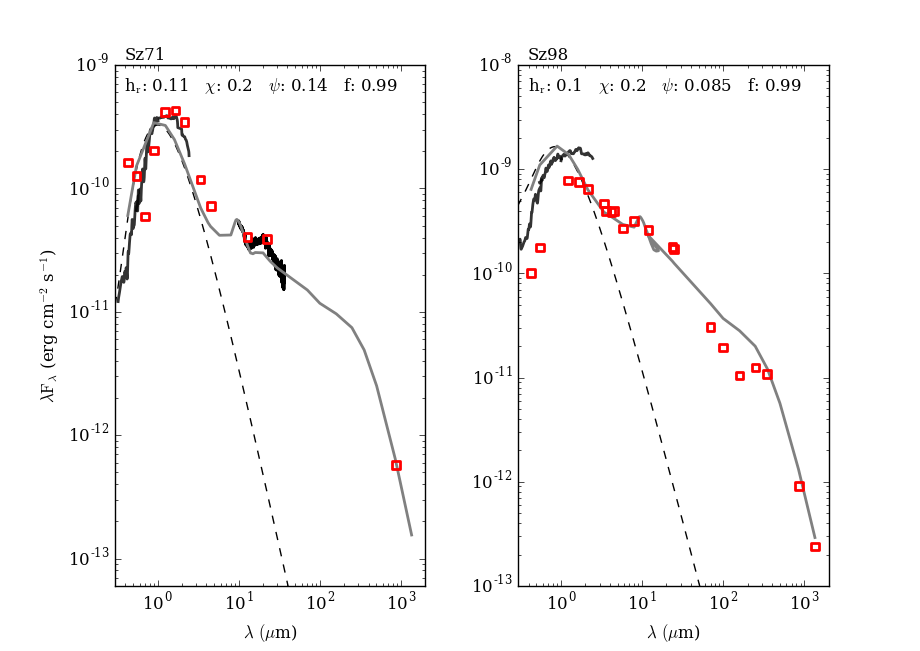}
	\caption{\texttt{DALI} SEDs and model parameters for the fiducial disk models. The preferred model is in gray, X-Shooter and IRS spectra are solid black lines, broadband continuum data are open red symbols, and the black dashed lines the input stellar photospheres.}
\end{center}
\end{figure*}

\renewcommand\thefigure{\thesection.\arabic{figure}}
\section{CN channel maps and emission line spectra for two bright, resolved sources}
\setcounter{figure}{0}
\label{app:chanmaps}

\begin{figure*}[ht]
	\begin{center}
		\includegraphics[width=0.95\textwidth]{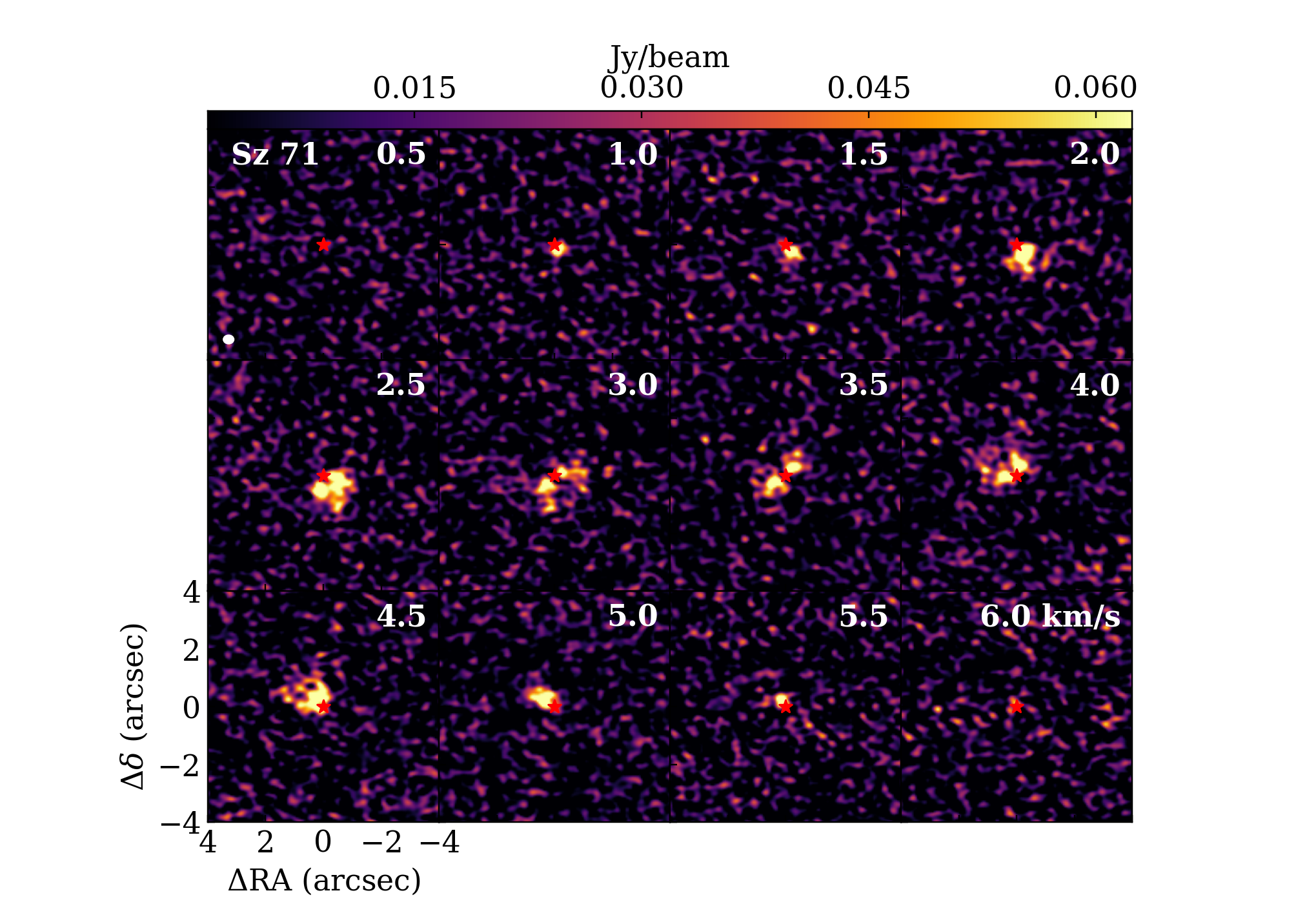}
		\caption{Channel map showing CN emission towards Sz 71, and relatively faint line wings indicative of an emission ring. The source location is marked with a red star; the contours are at the 3$\sigma$ and 5$\sigma$ levels.}
		\label{fig:sz71_chanmap}
	\end{center}
\end{figure*}

\begin{figure*}[hb]
	\begin{center}
		\includegraphics[width=0.95\textwidth]{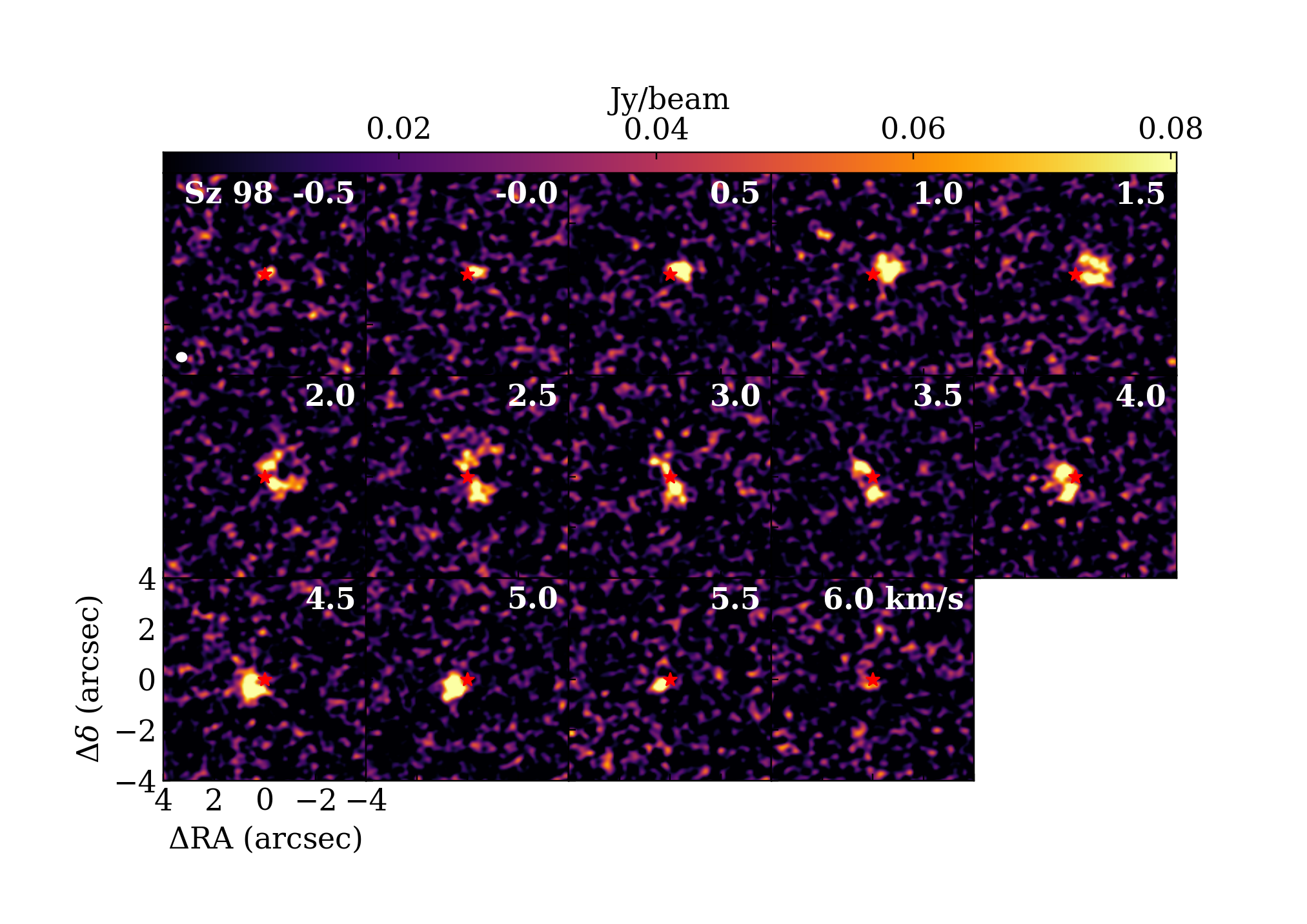}
		\caption{Channel map showing CN emission towards Sz 98, and relatively faint line wings indicative of an emission ring. The source location is marked with a red star; the contours are at the 3$\sigma$ and 5$\sigma$ levels.}
		\label{fig:sz98_chanmap}
	\end{center}
\end{figure*}

\begin{figure*}[ht]
\begin{center}
	\begin{subfigure}[a]{1.\textwidth}
		\includegraphics[width=17cm]{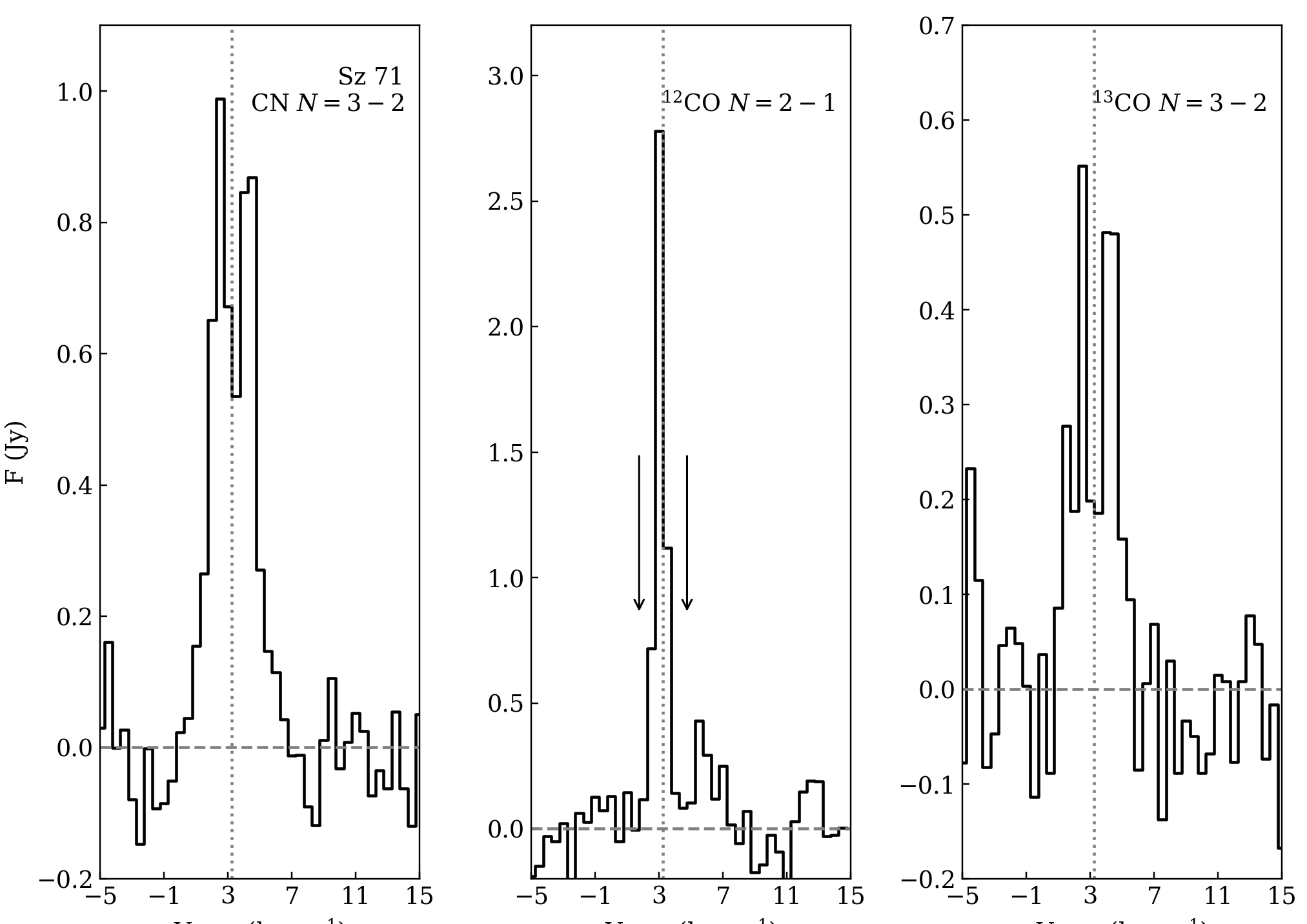}
		\label{fig:2a}
	\end{subfigure}

	\begin{subfigure}[b]{1.\textwidth}
		\includegraphics[width=17cm]{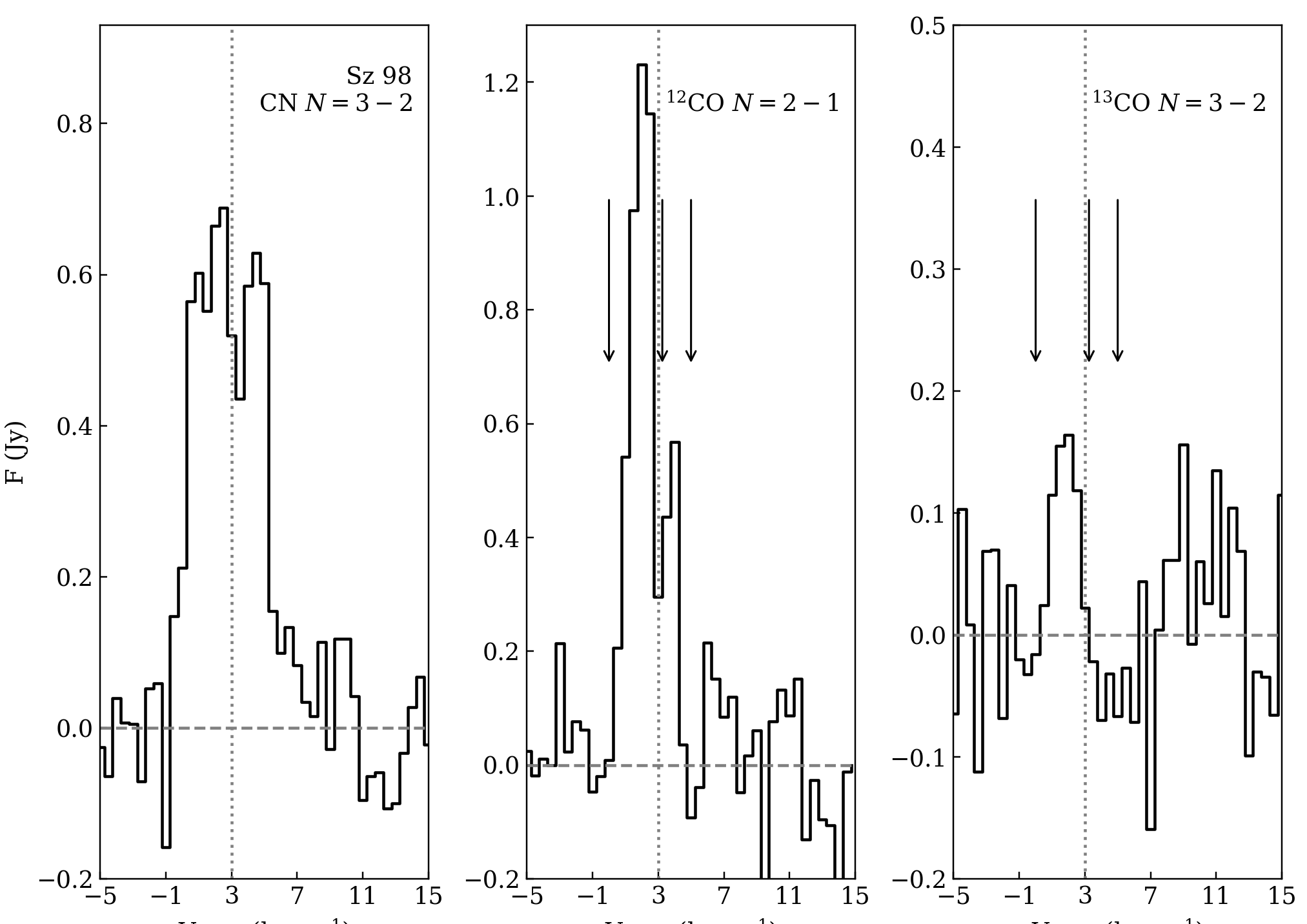}
		\label{fig:2b}
	\end{subfigure}
	\caption{Spectra of CN (focusing on the brightest $N=3-2,\,J=7/2-5/2$ components) and the main CO isotopologues for Sz 71 (top) and Sz 98 (bottom). Arrows show the velocities of foreground cloud material (partially) absorbing the $^{12}$CO line. Vertical dotted lines indicate the source velocity. The scale of the y-axis varies between panels. All spectra were extracted in an aperture overlapping with the detection in the moment-zero map (for $^{12}$CO), or in a $2''$ aperture on the source location (for CN and $^{13}$CO)}
\label{fig:2}
\end{center}
\end{figure*}

\section{CN fluxes and upper limits for the Lupus sample}
\label{app:fluxtable}
\longtab[1]{
	\begin{longtable}{l l l l}
	\caption{\label{tab:all_cn} CN $N=3-2,\,J=7/2-5/2$ fluxes, errors, and apertures}\\
	\hline \hline
	Source & $F_{\rm{CN}\,N=3-2}$ & $E_{\rm{CN}\,N=3-2}$ & Aperture \\
	 & [Jy\,km\,s$^{-1}$] & [Jy\,km\,s$^{-1}$] & [arcsec] \\
	\hline
	\endfirsthead
	\caption{continued.}\\
	\hline \hline
	Source & $F_{\rm{CN}\,N=3-2}$ & $E_{\rm{CN}\,N=3-2}$ & Aperture \\
	 & [Jy\,km\,s$^{-1}$] & [Jy\,km\,s$^{-1}$] & [arcsec] \\
	\hline
	\endhead
	\hline
	\endfoot
	Sz 65 & $250$ & $31$ & $0.58$ \\
	Sz 66 & $57$ & $14$ & $0.29$ \\
	J15430131-3409153 & $< 635$ & $...$ & $0.35$ \\
	J15430227-3444059 & $< 133$ & $...$ & $0.35$ \\
	J15445789-3423392 & $< 180$ & $...$ & $0.35$ \\
	J15450634-3417378 & $< 139$ & $...$ & $0.35$ \\
	J15450887-3417333 & $245$ & $51$ & $0.58$ \\
	Sz 68 & $191$ & $44$ & $0.58$ \\
	Sz 69 & $< 344$ & $...$ & $0.35$ \\
	Sz 71 & $2353$ & $95$ & $1.45$ \\
	Sz 72 & $< 95$ & $...$ & $0.35$ \\
	Sz 73 & $< 121$ & $...$ & $0.35$ \\
	Sz 74 & $< 111$ & $...$ & $0.35$ \\
	Sz 75 & $< 69$ & $...$ & $0.35$ \\
	Sz 76 & $949$ & $202$ & $1.62$ \\
	Sz 77 & $< 124$ & $...$ & $0.35$ \\
	Sz 81A & $< 166$ & $...$ & $0.35$ \\
	J15560210-3655282 & $890$ & $238$ & $1.16$ \\
	IM Lup \footnote{CN $N=2-1,\,J=5/2-3/2$ flux scaled to the CN $N=3-2,\,J=7/2-5/2$ transition.} 
		 & $3289$ & $130$ & \\
	Sz 83 & $1630$ & $135$ & $1.45$ \\
	Sz 84 & $626$ & $68$ & $1.16$ \\
	Sz 129 & $724$ & $61$ & $1.16$ \\
	J15592523-4235066 & $< 109$ & $...$ & $0.35$ \\
	RY Lup & $1624$ & $100$ & $1.16$ \\
	J16000060-4221567 & $< 106$ & $...$ & $0.35$ \\
	J16000236-4222145 & $1983$ & $114$ & $2.03$ \\
	J16002612-4153553 & $< 108$ & $...$ & $0.35$ \\
	Sz 130 & $< 140$ & $...$ & $0.35$ \\
	MY Lup & $1700$ & $123$ & $1.45$ \\
	Sz 131 & $< 89$ & $...$ & $0.35$ \\
	J16011549-4152351 & $1131$ & $86$ & $1.16$ \\
	EX Lup & $< 63$ & $...$ & $0.35$ \\
	Sz 133 & $2655$ & $112$ & $1.16$ \\
	Sz 88A & $< 243$ & $...$ & $0.35$ \\
	Sz 88B & $< 95$ & $...$ & $0.35$ \\
	J16070384-3911113 & $430$ & $67$ & $1.16$ \\
	J16070854-3914075 & $1719$ & $91$ & $1.16$ \\
	Sz 90 & $273$ & $42$ & $0.58$ \\
	Sz 91 & $...$ & $...$ &  \\
	J16073773-3921388 & $50$ & $13$ & $0.29$ \\
	Sz 95 & $< 94$ & $...$ & $0.35$ \\
	J16075475-3915446 & $< 388$ & $...$ & $0.35$ \\
	J16080017-3902595 & $< 183$ & $...$ & $0.35$ \\
	J16080175-3912316 & $< 202$ & $...$ & $0.35$ \\
	Sz 96 & $< 145$ & $...$ & $0.35$ \\
	J16081497-3857145 & $160$ & $37$ & $0.58$ \\
	Sz 97 & $< 152$ & $...$ & $0.35$ \\
	Sz 98 & $3494$ & $131$ & $1.74$ \\
	Sz 99 & $< 149$ & $...$ & $0.35$ \\
	Sz 100 & $350$ & $45$ & $0.87$ \\
	J160828.1-391310 & $< 235$ & $...$ & $0.35$ \\
	Sz 102 & $< 83$ & $...$ & $0.35$ \\
	Sz 103 & $< 258$ & $...$ & $0.35$ \\
	J16083070-3828268 & $7467$ & $134$ & $2.03$ \\
	Sz 104 & $< 105$ & $...$ & $0.35$ \\
	J160831.1-385600 & $< 132$ & $...$ & $0.35$ \\
	V856 Sco & $< 205$ & $...$ & $0.35$ \\
	V1094 Sco \footnote{Partly resolved-out and continuum-shielded}
				& $3430$ & $475$ & $5.34$ \\
	Sz 106 & $< 152$ & $...$ & $0.35$ \\
	Sz 108B & $115$ & $32$ & $0.58$ \\
	J16084940-3905393 & $< 98$ & $...$ & $0.35$ \\
	V1192 Sco & $< 112$ & $...$ & $0.35$ \\
	Sz 110 & $< 95$ & $...$ & $0.35$ \\
	J16085324-3914401 & $119$ & $26$ & $0.46$ \\
	J16085373-3914367 & $182$ & $38$ & $0.58$ \\
	Sz 111 & $1881$ & $106$ & $2.03$ \\
	J16085529-3848481 & $< 190$ & $...$ & $0.35$ \\
	Sz 112 & $< 82$ & $...$ & $0.35$ \\
	Sz 113 & $140$ & $33$ & $0.58$ \\
	J16085828-3907355 & $< 353$ & $...$ & $0.35$ \\
	J16085834-3907491 & $< 126$ & $...$ & $0.35$ \\
	J16090141-3925119 & $805$ & $47$ & $0.87$ \\
	Sz 114 & $1158$ & $55$ & $1.16$ \\
	Sz 115 & $< 150$ & $...$ & $0.35$ \\
	J16091644-3904438 & $< 272$ & $...$ & $0.35$ \\
	J16092032-3904015 & $< 277$ & $...$ & $0.35$ \\
	J16092317-3904074 & $< 136$ & $...$ & $0.35$ \\
	J16092697-3836269 & $< 268$ & $...$ & $0.35$ \\
	J160934.2-391513 & $< 135$ & $...$ & $0.35$ \\
	J16093928-3904316 & $< 281$ & $...$ & $0.35$ \\
	Sz 117 & $< 153$ & $...$ & $0.35$ \\
	Sz 118 & $295$ & $65$ & $0.87$ \\
	J16095628-3859518 & $< 95$ & $...$ & $0.35$ \\
	J16100133-3906449 & $< 132$ & $...$ & $0.35$ \\
	J16101307-3846165 & $< 106$ & $...$ & $0.35$ \\
	J16101857-3836125 & $< 107$ & $...$ & $0.35$ \\
	J16101984-3836065 & $< 89$ & $...$ & $0.35$ \\
	J16102741-3902299 & $< 130$ & $...$ & $0.35$ \\
	J16102955-3922144 & $787$ & $48$ & $0.87$ \\
	Sz 123B & $< 102$ & $...$ & $0.35$ \\
	Sz 123A & $241$ & $41$ & $0.64$ \\
	J16115979-3823383 & $< 95$ & $...$ & $0.35$ \\
	J16120445-3809589 & $< 219$ & $...$ & $0.35$ \\
	J16124373-3815031 & $1015$ & $82$ & $0.99$ \\
	J16134410-3736462 & $< 149$ & $...$ & $0.35$ \\
	\hline
	\end{longtable}
}

\end{appendix}

\end{document}